\documentclass[prd,preprint,tightenlines,floatfix,showpacs,date,preprintnumbers,nofootinbib,eqsecnum]{revtex4}
 \usepackage[dvips,final]{graphicx}
  \usepackage{bm}
   \usepackage{amsmath}
    \usepackage{amssymb}
     \usepackage{pifont}
\usepackage{color}
 
  \newcommand{\BluTn}[1]{\textcolor{blue}{#1}}
   \newcommand{\RedTn}[1]{\textcolor{red}{#1}}
\newcommand{\be}{\begin{equation}}\newcommand{\ee}{\end{equation}}%
\newcommand{\bd}{\begin{displaymath}}\newcommand{\ed}{\end{displaymath}}
\newcommand{\bit}{\begin{itemize}}                                        
 \newcommand{\eit}{\end{itemize}}                                         
\newcommand{\ben}{\begin{enumerate}}                                      
 \newcommand{\een}{\end{enumerate}}                                       
\newcommand{\baa}{\begin{array}{lll}}                                     
 \newcommand{\eaa}{\end{array}}                                           
\newcommand{\ba}{\begin{eqnarray}}                                        
 \newcommand{\ea}{\end{eqnarray}}                                         
\newcommand{\la}{\label}                                                  
\newcommand{\Ds}{\displaystyle}                                           
\newcommand{\gev}[1]{\relax\ifmmode{\text{GeV}^{#1}}                      
                     \else{GeV$^{#1}${ }}\fi}                             
\def\MSbar{\relax\ifmmode\overline                                        
            {\rm MS}\else{$\overline{\rm MS}${ }}\fi}                     
\def\as{\relax\ifmmode \alpha_s\else{$ \alpha_s${ }}\fi}                  
\def\abar{\relax\ifmmode{\bar{a}}\else{$\bar{a}${ }}\fi}                  
\newlength{\tabcolf} 
\newlength{\tabcols} 
\newlength{\tabcolt} 

\begin{document}
\date{\today}
~~~~~~~~~~~~~~~~~~~~~~~~~~~~~~~~~~~~~~~~~~~~~~~~~~~~~~~~~~~~~~~
\preprint{RUB-TPII-05/09}
\title{Transition form factors of the pion in light-cone QCD
       sum rules with next-to-next-to-leading order contributions}
\author{S.~V.~Mikhailov}%
 \email{mikhs@phys.jinr.ru}
\affiliation{%
  Bogoliubov Laboratory of Theoretical Physics,
  JINR, 141980 Dubna, Moscow Region, Russia}%
\author{N.~G.~Stefanis}
 \email{stefanis@tp2.ruhr-uni-bochum.de}
\affiliation{%
  Institut f\"ur Theoretische Physik II,
  Ruhr-Universit\"at Bochum,
  D-44780 Bochum, Germany}%
\affiliation{%
  Bogoliubov Laboratory of Theoretical Physics,
  JINR, 141980 Dubna, Moscow Region, Russia}%
\vspace {10mm}
\begin{abstract}
The transition pion-photon form factor is studied within the
framework of Light-Cone QCD Sum Rules.
The spectral density for the next-to-leading order corrections
is calculated for any Gegenbauer harmonic.
At the level of the next-to-next-to-leading (NNLO) radiative
corrections, only that part of the hard-scattering amplitude is
included that is proportional to the $\beta$-function,
taking into account the leading zeroth-order harmonic.
The relative size of the NNLO contribution in the prediction for the
form factor $F^{\gamma^{*}\gamma\pi}(Q^2)$ has been analyzed, making
use of the BLM scale-setting procedure.
In addition, predictions for the form factor $F^{\gamma^{*}\rho\pi}$
are obtained that turn out to be sensitive to the endpoint
behavior of the pion distribution amplitude, thus providing in
connection with experimental data an additional adjudicator for
the pion distribution amplitude.
In a note added, we comment on the preliminary high-$Q^2$ BaBar
data on $F^{\gamma^{*}\gamma\pi}$ arguing that the significant growth
of the form factor between 10 and 40 GeV$^2$ cannot be explained in
terms of higher-order perturbative corrections at the NNLO.
\end{abstract}
\vspace {2mm}

\maketitle
\newpage

\section{I\lowercase{ntroduction}}
\label{sec:intro}
Although higher-order calculations in QCD perturbation theory have
already a long history, little is known about exclusive processes
at the next-to-leading order (NLO) level
\cite{DaCh81,Br83,KMR86,MNP99a,SchmYa99,BMS02},
and beyond \cite{MMP02,MNP01a}, because these are quite complex in
detail.
In view of more and more high-precision experimental data for a
variety of hadronic processes becoming gradually available, the
importance of such higher-order calculations exceeds the pure
theoretical interest and acquires phenomenological relevance.
In particular, processes with two photons in the initial state, one
far off-shell and the other quasi real,
$$
  \gamma^* + \gamma \rightarrow \pi^0 \ ,
$$
provide a useful tool to access (after their fusion) the partonic
structure of the produced hadronic states, e.g., pseudoscalar mesons.

Experimentally, the photon-to-pion transition form factor
within this class of two-photon processes has been measured by the CLEO
Collaboration \cite{CLEO98} with high precision and extending the range
of $Q^2$ up to 9 GeV$^2$, as compared to the previous low-momentum
CELLO data \cite{CELLO91}.
Theoretically, this high precision allows one to test models and
fundamental quantities, like the pion distribution amplitude (DA), the
applicability of QCD factorization, etc.---see
\cite{JKR96,KR96,RR96,Kho99,SchmYa99,SSK99,SSK00,DKV01,BMS01,%
BMS02,BMS03,Ag04,Ag05b,BMS05lat,CN07,BroArr07,BMPS07,Ste08,%
Guo-Liu08}
and references cited therein.
Moreover, one can determine \cite{Ste08} a compatibility region between
the CLEO data and constraints derived from lattice simulations on the
second moment of the pion DA \cite{Lat06,Lat07}.
This information can then be used to extract a range of values of the
fourth moment of the pion DA that would simultaneously fulfil both
constraints (CLEO and lattice).
This prediction \cite{Ste08} can provide a guide for the determination
of this moment on the lattice, a task that has not been accomplished
yet.

For two highly virtual photons, perturbative QCD works well because
factorization at some factorization scale $\mu_{\rm F}^{2}$ applies,
so that the process can be cast into the form of a convolution
\begin{equation}
  F^{\gamma^{*}\gamma^{*}\pi}(Q^{2},q^{2})
=
 C\left(Q^{2},q^{2},\mu_{\rm F}^{2},x\right)
\otimes
 \varphi_{\pi}\left(x,\mu_{\rm F}^{2}\right)
 + \mathcal{O}(Q^{-4}) \ ,
\label{eq:convolution}
\end{equation}
which contains a hard part $C$, calculable within perturbation theory,
and a wave-function part $\varphi_{\pi}$ that is the (leading)
twist-two pion distribution amplitude \cite{Rad77} and has to be
modeled within some nonperturbative framework (or be extracted from
experiment).
Here, the omitted twist-four contribution represents subleading terms
in the operator product expansion (OPE), which are suppressed by
inverse powers of the photon virtualities.

\begin{figure}[ht]
\centerline{\includegraphics[width=0.6\textwidth]{
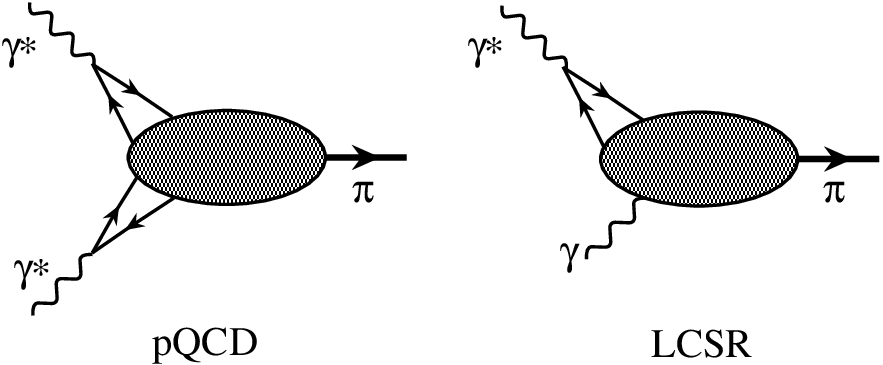}}
\vspace*{-0mm}
\caption{\footnotesize Pion-photon transition form factor in
         perturbative QCD (left) and using Light Cone Sum Rules (LCSR)
         (right).
         The right graphics illustrates the situation when one
         of the two photons is real and perturbation theory becomes
         inapplicable, because the hadronic content of the photon
         starts to be relevant.}
\label{fig:blobs}
\end{figure}

To be more precise, consider the hard process of two colliding
photons producing a single pion,
$\gamma^{*}(q_{1})\gamma^{*}(q_{2})\to\pi^{0}(p)$,
which is defined by the following matrix element \cite{Kho99}
\begin{equation}
  \int d^{4}\bm{z} e^{-iq_{1}\cdot z}
  \langle
         \pi^0 (p)\mid T\{j_\mu(z) j_\nu(0)\}\mid 0
  \rangle
=
  i\epsilon_{\mu\nu\alpha\beta}
  q_{1}^{\alpha} q_{2}^{\beta} F^{\gamma^{*}\gamma^{*}\pi}(Q^2,q^2)\, ,
\label{eq:matrix-elem}
\end{equation}
where
$Q^2=-q_1^2$, $q^2=-q_2^2$ denote the virtualities of the photons,
$\pi^{0}(p)$ is the pion state with the momentum $p=q_1+q_2$,
and
$
 j_{\mu}
=
 (\frac{2}{3}\bar{u}\gamma_{\mu}u - \frac{1}{3}\bar{d}\gamma_{\mu}d)
$
is the quark electromagnetic current.
This process is illustrated graphically in the left panel of
Fig.\ \ref{fig:blobs} and has been examined theoretically, for
instance, in \cite{LB80,DaCh81,Br83,KMR86,DKV01}.

If both virtualities, $Q^2$ and $q^2$, are sufficiently large, the
$T$-product of the currents can be expanded near the light cone
($z^{2}=0$) by virtue of the OPE to obtain the well-known leading-order
expression for the convolution in Eq. (\ref{eq:convolution})
\begin{equation}
  F^{\gamma^{*}\gamma^{*}\pi}(Q^{2},q^{2})
=
  N_{\rm T}
  \int_{0}^{1} dx
  \frac{1}{Q^{2}\bar{x} + q^{2}x}~\varphi_{\pi}(x)\, ,
\label{eq:leading-FF-term}
\end{equation}
where we have used the abbreviations
$\bar{x}\equiv 1-x$
and
\begin{equation}
  N_{\rm T}
\equiv
  (e^{2}_{u}-e^{2}_{d})\sqrt{2}\, f_{\pi}
=
  \frac{\sqrt{2}}{3}f_{\pi} \ .
\end{equation}

In contrast, the kinematics probed in the CELLO \cite{CELLO91} and
the CLEO \cite{CLEO98} experiments involves a quasi-real
photon with $q^2 \to 0$.
At such a low virtuality, the hadronic content of the quasi-real
photon, i.e., its long-distance structure \cite{Kho99} becomes
important
(see the diagram in the right panel of Fig.\ \ref{fig:blobs}),
thus preventing a straightforward QCD calculation \cite{RR96} of the
form factor
$
 F^{\gamma^{*}\gamma\pi}(Q^2, q^2 \to 0)
\equiv
 F^{\gamma^{*}\gamma\pi}(Q^2)
$ on the ground of factorization.
The method of Light-Cone QCD Sum rules (LCSR for short) allows one to
avoid this problem by providing the means of performing all QCD
calculations at sufficiently large $q^2$ ($\gamma^*$) and then use a
dispersion relation to ``approach'' the mass-shell photon
($\gamma$) with zero virtuality.
This calculational scheme, which can accommodate the large-distance
properties of the photon, i.e., its hadronic content, was proposed by
Khodjamirian in \cite{Kho99} and the form factor
$F^{\gamma^*\gamma \pi}(Q^2)$
was calculated at the LO level of the LCSR including also twist-four
contributions.

The core ingredient of the LCSR approach is the spectral density,
which provides a powerful tool for a quantitative description of
hadronic processes in QCD in terms of a dispersion relation:
\begin{equation}
  F^{\gamma^{*}\gamma\pi}(Q^2,q^2)
=
  \int\limits_{0}^\infty ds~ \frac{\rho^{\rm phen}(Q^2,s)}{s+q^2}\, .
\label{eq:Dispersion}
\end{equation}
[Note that here the label ``phen'' abbreviates phenomenological].
Effects due to the long-distance dynamics of the
$\gamma^*\gamma\rightarrow \pi^0$ process are partly contained in the
form factor $F^{\gamma^{*}\rho\pi}(Q^2)$ (which can be obtained by
means of quark-hadron duality) and also in the $\pi$
distribution amplitudes of different twists
\cite{ER80,ER80tmf,LB79,LB80}.
Within the LCSR approach the form factor $F^{\gamma^{*}\rho\pi}(Q^2)$
appears inevitably because one assumes that the spectral density,
entering the dispersion relation, can be approximated by the ground
states of vector mesons \cite{Kho99,BMS04kg}, like the $\rho$ and the
$\omega$.\footnote{For the sake of simplicity, one sets the masses of
the two vector mesons equal and appeals to isospin symmetry to treat
both particles in terms of a combined effective resonance.}

At this point, two important remarks are in order.
(i) Contrary to previous calculations, e.g., in \cite{SchmYa99,BMS02},
we will not adopt a zero-width approximation here, but use instead a
more realistic Breit-Wigner ansatz for the effective resonances
(see Sec.\ \ref{sec:data-comp}).
(ii) The scaled form factor
$Q^{4}F^{\gamma^{*}\rho\pi}(Q^2)$, obtained in the framework of LCSRs,
depends at large $Q^2$ mainly on the \emph{differential} pion
characteristic
$\displaystyle
 \frac{d}{dx} \varphi_{\pi}(x)|_{x=\epsilon}
\, ,
$
with
$\displaystyle\epsilon \sim \frac{s_0}{Q^2}\ll 1 $
being a small neighborhood around the origin, where $s_0$ is the
duality interval entering the model for $\rho^{\rm phen}$.
This feature appears to be opposite to the case of the
$Q^{2}F^{\gamma^{*}\gamma\pi}(Q^2)$ (scaled) form factor, that depends
mainly on (though it is not directly proportional to)
the inverse moment \cite{BMS05lat}
$
 \langle x^{-1}\rangle_{\pi}
=
 \int^{1}_{0} \varphi_{\pi}(x;\mu^{2}){x}^{-1}dx
$,
cf.\ Eq.\ (\ref{eq:leading-FF-term}) evaluated at the scale
$q^2 \to 0$, because the latter is an \emph{integral}
characteristic of the pion DA \cite{BMS01, BMS03}.
Hence, $Q^{4}F^{\gamma^{*}\gamma\rho\pi}(Q^2)$ can provide
complementary information on the pion DA and help
discriminate among various proposed pion DA models.

The structure of the paper is the following.
In the next section, we recall the formalism of LCSRs for the form
factors
$F^{\gamma^{*}\gamma\pi}(Q^2)$, $F^{\gamma^{*}\rho\pi}(Q^2)$
and construct the spectral density in a systematic way.
This calculation is extended beyond the LO in Sec.\
\ref{sec:struc-spec-dens} and an explicit expression for the
spectral density at the NLO for any index $n$---the latter indicating
the order of the expansion in Gegenbauer harmonics---is derived.
The further extension to the NNLO is also given in this section.
Actually, we include only the $\beta_0$-proportional contributions
that can be obtained from the corresponding terms of the
hard-scattering amplitudes and denote it by NNLO$_{\beta}$.
The effects of the NNLO contributions are discussed in Sec.\
\ref{sec:FF-results} in connection with the BLM prescription
(and its modifications \cite{BPSS04}).
Our predictions for the form factors
$F^{\gamma^{*}\gamma\pi}(Q^2)$, $F^{\gamma^{*}\rho\pi}(Q^2)$ are
presented in Sec.\ \ref{sec:data-comp}, where we also provide a
comparison of $F^{\gamma^{*}\gamma\pi}(Q^2)$ with the experimental
data.
Section \ref{sec:concl} contains our conclusions emphasizing our
main results.
Important technical details are provided in three dedicated appendices.
In a Note Added, we point out that the new BaBar data on
$F^{\gamma^{*}\gamma\pi}$, which show a significant growth of the
form factor beyond 10~GeV$^2$, cannot be described within the QCD
convolution approach.

\section{$F^{\gamma^{*}\gamma\pi}(Q^2)$, $F^{\gamma^{*}\rho\pi}(Q^2)$
         in light-cone sum rules. Formalism}
\label{sec:LCSR-formalism}
Here we present the theoretical description of the transition form
factor $F^{\gamma^{*}\gamma\pi}(Q^{2})$ of the exclusive process
$\gamma^{*}\gamma\to\pi^{0}$, employing the framework of
light-cone sum rules \cite{BBK89,BF89,BH94,Kho99} beyond
the next-to-leading-order (NLO) of perturbative QCD.\footnote{For
the sake of clarity, we use the following notation for the form
factors: The associated reaction is denoted by superscripts,
whereas the calculational context, e.g., QCD, is marked by
subscripts.
Note that our notation differs from abbreviated notations used in
the cited works.}
An integral part of this sort of calculation is the form
factor $F^{\gamma^{*}\rho\pi}(Q^{2})$, describing the transition
$\gamma^{*}\rho\to\pi$, which will, therefore, be computed
in parallel.

\subsection{Factorization}
\label{subsec:fact}
The calculation of $F^{\gamma^{*}\gamma\pi}(Q^2)$ proceeds through the
following main steps:
(i) First, the form factor
$F_\text{QCD}^{\gamma^*\gamma^*\pi}(Q^2,q^2)$
at large Euclidean virtualities of the photons,
$Q^2, q^2\geq 1$~GeV${}^2$ is calculated.
(ii) Then, an appropriate realistic model for the spectral density at
low $s$, based, for instance, on quark-hadron duality, is constructed.
(iii) Finally, the dispersion relation for the form factor
$F^{\gamma \gamma^*\pi}(Q^2,q^2)$ is exploited.

Applying the factorization theorems \cite{ER80,ER80tmf,LB79,LB80},
the form factor $F_\text{QCD}^{\gamma^*\gamma^*\pi}(Q^2,q^2)$
can be cast in the form
\ba
   F_\text{QCD}^{\gamma^*\gamma^* \pi}(Q^2,q^2)
= N_{\rm T}  T\otimes
  \varphi_{\pi} + \mbox{higher-twist contributions}\ .
\label{eq:factor}
\ea
The hard-scattering amplitude $T$ for this process, written below
in the square brackets,
\ba
   F_\text{QCD}^{\gamma^*\gamma^* \pi}(Q^2,q^2)
=
   N_{\rm T}
   \left[
         T_0(Q^2,q^2;x)+ \right.
& \left. \!\!\!\!\!\! \!\!\!\!\!\!\!\!\!\!\!\!\!
   a_s^1~T_1(Q^2,q^2;\mu^2_F;x)\right.
& \nonumber \\
+ & \!\!\!\!\!\!\!\!\!\! \left.
   a_s^2~T_2(Q^2,q^2;\mu^2_{\rm F};\mu_\text{R}^2;x)\right.
&  \nonumber \\
+ & \left.  \ldots \right.
&\left.
   \!\!\!\!\!\!\! \right]
\otimes
    \varphi_{\pi}^{(2)}(x;\mu^2_{\rm F})\Big|_{\mu^2_{\rm F}=Q^2}
\nonumber \\
    +& \mbox{higher-twist contributions,}&
\label{eq:PTformfactor}
\ea
\ba
 a_s
=
 \frac{\alpha_s(\mu_\text{R}^2)}{4\pi},\qquad
 \otimes \equiv \int_0^1 dx,
\label{eq:couplant}
\ea
is calculable within perturbative QCD.
The symbols $\mu_\text{R}$ and $\mu_\text{F}$ denote, respectively,
the scale of the renormalization of the theory and the factorization
scale of the process.
The pion DA $\varphi^{(2)}_\pi(x;\mu_\text{F}^2)$ of twist two,
entering the convolution with the hard-scattering amplitude,
is inaccessible to perturbative QCD and demands the application of
nonperturbative methods (see Section \ref{subsec:nonperturbative}).
We quote the well-known result for
$F_\text{QCD}^{\gamma^*\gamma^* \pi}$
at LO in $a_s$ that also includes
the twist-four contribution \cite{Kho99}, viz.,
\ba
  F_\text{LO QCD}^{\gamma^*\gamma^*\pi}(Q^2,q^2)
&=&
  N_{\rm T}
           \left\{
                  T_0(Q^2,q^2;x) \otimes \varphi^{(2)}_\pi(x)
                  - \left[T_0(Q^2,q^2;x)\right]^2
                  \otimes \varphi^{(4)}_\pi(x)
           \right\}
\label{eq:FF-LO}
\ea
with
\ba
 ~T_0(Q^2,q^2)
&=&
  \frac1{2}\frac{1}{Q^2\bar{x}+q^2x} + (x\to \bar{x})
\label{eq:T0}
\ea
and where $\varphi^{(4)}_\pi(x)$ is a naturally appearing combination
of twist-four pion DAs (for more details, see \cite{BraM07} and
Sec.\ \ref{subsec:nonperturbative}).
An explicit expression for $T_1$ has been obtained in
\cite{DaCh81,Br83,KMR86}.
More recently \cite{MMP02}, the general structure of $T_2$ was
investigated and its $\beta$--part, $b_0 \cdot T_{\beta}$, was
calculated.
For convenience, both amplitudes are included in
Appendix~\ref{app:Tstructure}.
In fact, the hard-scattering amplitudes
$T_{0}, T_{1}, T_{2}$
also determine the spectral densities
$\rho^{(0)}, \rho^{(1)}, \rho^{(2)}$,
as one can see from the following generic expression
\be
  \rho(Q^2,s)
=
  \frac{\mathbf{Im}}{\pi}
  \left[\left(T\otimes \varphi_{\pi}\right)(Q^2,-s)
  \right]\,, ~s\geq 0 \ .
\label{eq:defrho}
\ee
The core issues of the hard-scattering amplitudes are listed below,
while characteristic Feynman graphs for each order of the
perturbative expansion are depicted for illustration in Fig.\
\ref{fig:graphs}:
\begin{figure}[ht]
\centerline{\includegraphics[width=0.7\textwidth]{
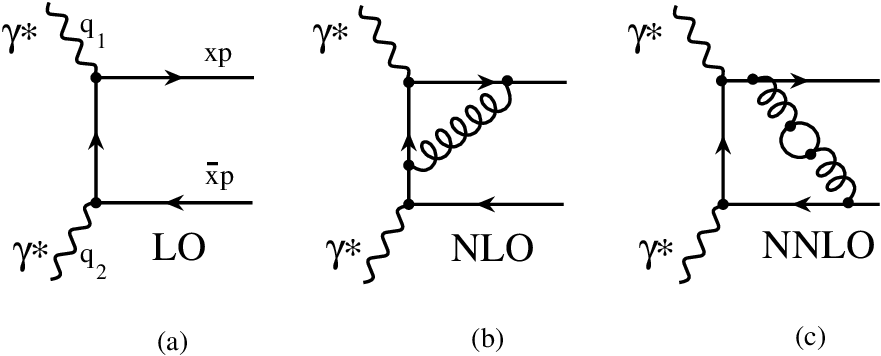}}
\vspace*{-3mm}
\caption{\footnotesize Typical Feynman graphs contributing at different
          orders of the perturbative expansion of the pion-photon
          transition form factor in QCD.
          (a) LO, (b) NLO, and (c) NNLO, with the momenta of the
          various particles being indicated explicitly in (a).
          Both colliding photons are considered to be highly virtual.}
\label{fig:graphs}
\end{figure}

\begin{enumerate}
\item
\textbf{LO} perturbative QCD: $T_0$, see Fig.\ \ref{fig:graphs}(a). \\
In this order of the expansion we have
$T_0(Q^2,q^2;x)\otimes\varphi(x;\mu_{\rm F}^{2})$
and the factorization scale cannot be determined uniquely.
\item
\textbf{NLO} perturbative QCD: $T_1$.
A typical diagram is shown in Fig.\ \ref{fig:graphs}(b). \\
Here the hard-scattering amplitude starts depending upon
the factorization scale $\mu_\text{\rm F}$.
In explicit terms this reads \\
$
 T_{1}\left(x;Q^2,q^2\right)\otimes\varphi\left(x\right)
=
  T_{0}(Q^2,q^2;y)\otimes\left\{ C_{\rm F} {\cal T}^{(1)}(y,x)
+
 {\rm Ln}(y) V^{(0)}(y,x)\right\} \otimes\varphi\left(x\right)
$ \\
and the $\mu_{\rm F}$-dependence enters via the logarithm
\begin{equation}
 {\rm Ln}(y)
\equiv
 \ln\left[\left(Q^2y+q^2\bar{y}\right)/\mu^2_\text{F}\right]\ ,
\label{eq:L-n}
\end{equation}
whereas the LO Efremov-Radyushkin-Brodsky-Lepage (ERBL) kernel
\cite{ER80,ERBL79,LB80} $V^{(0)}$ is given in the next section with
more details provided in Appendix \ref{app:Vstructure}.
As regards the term ${\cal T}^{(1)}(y,x)$, defined in
(\ref{Str-T1-log}), it contains contributions from the partial kernel
$V^{b}$ (which enters $V^{(0)}$ and is generated by
contracting the diagram in Fig.\ \ref{fig:graphs}(b)) and also the
kernel $g$ stemming from $V^{(1)}$ and defined in (\ref{eq:Str-T1}).
\item
\textbf{NNLO} perturbative QCD: $T_2$ with an example depicted
graphically in Fig.\ \ref{fig:graphs}(c). \\
In this case, things are more complicated.
First of all, we reiterate that only the $\beta_0$-proportional
contributions to $T_2$ (termed $T_{\beta}$) will be considered,
making use of the calculations performed in \cite{MMP02}.
This is, because that part can provide the right size of the whole NNLO
contribution using the scheme-independent BLM scale-setting procedure
to eliminate the appearance of the $\beta$-function in the perturbative
series.
To discuss the structure of $T_{\beta}$, let us first present it
explicitly (postponing details to the appropriate sections to follow):
\begin{eqnarray}
  T_{\beta}\otimes\varphi
= &&\!\!\!
          \ln\left(\frac{\mu_{\rm R}^2}{\mu_{\rm F}^2}\right)T_{1}
\otimes
        \varphi + T_{0}
\otimes
        \left\{
                 C_{\rm F}{\cal T}^{(2)}_{\beta}
               + C_{\rm F} {\rm Ln}(y)
                 \left[\left(V^{(1)}_{\beta}\right)_+-{\cal T}^{(1)}
                 \right] \right. \nonumber \\
&&  \left.
        - \frac{1}{2}{\rm Ln}^2(y)V^{(0)}
        \right\}\otimes\varphi \, .
\label{eq:T-beta}
\end{eqnarray}
Second, as one sees from this expression, the hard-scattering amplitude
depends explicitly also on the renormalization scale, $\mu^2_{\rm R} $,
owing to the renormalization of the strong running coupling.
As a result, additional logarithms of the form
$\ln\left(\mu_{\rm F}^{2}/\mu_{\rm R}^{2}\right)$ appear.
These terms are controlled by the renormalization-group (RG) equation
and can be resummed by applying the BLM procedure \cite{BLM83}
which amounts to a rescaling of the argument of the running coupling
to another value.
This will be discussed later in more detail.
Third, the logarithms ${\rm Ln}(y)$ and ${\rm Ln}^{2}(y)$, which
bear the $\mu_{\rm F}^{2}$-dependence [cf.\ Eq.\ (\ref{eq:L-n})],
are accompanied by elements of the kernel $V^{(1)}$ and the kernel
$V^{(0)}$ governing the NLO and LO evolution, respectively.
\end{enumerate}

\subsection{Construction of the spectral density}
\label{subsec:constr-spectr-dens}
We continue here with the systematic construction of the spectral
density.
In doing so, we will make use of another set of variables,
$(x,Q^2)$, instead of the usual set $(s,Q^2)$.
In the LO approximation, $\rho^{(0)}(x)$ follows from the definition
(\ref{eq:defrho}) upon inserting in Eq.\ (\ref{eq:T0}) the explicit
expression for $T_0(Q^2,q^2;x)$ \cite{Kho99,SchmYa99}.
Then, we obtain
\begin{subequations}
 \label{eq:rho0}
  \begin{eqnarray}
  \rho^{(0)}(Q^2,s)
&=&
  \frac{\mathbf{Im}}{\pi}
  \Big[
       T_0(Q^2,-s)\otimes \varphi^{(2)}_{\pi}+\mbox{tw-4}
  \Big]
\nonumber \\
&\equiv&
  \frac{1}{Q^2+s} \bar{\rho}(x,Q^2) \bm{\Big|_{x=\frac{Q^2}{Q^2+s}}}
\label{def:rhobar}
\\
&=&
  \frac{1}{Q^2+s}
  \left(
        \varphi^{(2)}_{\pi}(x)+\frac{x}{Q^2}\frac{d}{dx}
        \varphi^{(4)}(x)
  \right)
         \Big|_{x=\frac{Q^2}{Q^2+s}}
\label{eq:rho-0}
\end{eqnarray}
 \end{subequations}
[cf.\ first item in Appendix \ref{app:Tstructure}].
One notices that $\rho^{(0)}$ is directly proportional to the pion DA
of leading twist two, $\varphi^{(2)}_\pi$, and the derivatives with
respect to $x$ of the twist-four contribution, $\varphi^{(4)}$.
This observation allows one to simplify the subsequent analysis by
introducing  the normalized spectral density $\bar{\rho}(x,Q^2)$ via
Eq.~(\ref{def:rhobar}).

Expanding $\varphi^{(2)}_\pi(x;\mu^2)$ in terms of the eigenfunctions
$\psi_n(x)$ of the one-loop ERBL evolution equation, which coincide
with the Gegenbauer polynomials $C^{3/2}_n(\xi)$, we find
\begin{equation}
  \varphi^{(2)}_\pi(x;\mu^2)
=
  \psi_0(x) +  \sum_{n=2,4,\ldots}a_n(\mu^2) \psi_n(x)\, ,
\quad
  ~\psi_n(x)
\equiv
  6x(1-x)C^{3/2}_n(2x-1)\, .
\label{eq:Gegenbauer}
\end{equation}
In this representation, all scale dependence is contained in the
coefficients $a_n(\mu^2)$ and is controlled by the ERBL equation.
For this reason, it is convenient to project the leading-twist part of
the spectral density $\rho(Q^2,s)$ on the same basis of eigenfunctions
$\{\psi_n\}$ and introduce the partial density $\rho_n$, which has
the form
\ba
  \rho_n(Q^2,x)
&=&
  \frac{x}{Q^2}
   \bar{\rho}_n(x) \, ,
\label{eq:defbarrho}
\ea
where $x=Q^2/(Q^2+s)$ and
\ba
&&
  \bar{\rho}_n(x)
=
    \bar{\rho}^{(0)}_n(x) + a_s^{1}~\bar{\rho}^{(1)}_n(x)
  + a_s^{2}~\bar{\rho}^{(2)}_n(x)+ \ldots \, .
\label{eq:rho-n-bar}
\ea
Therefore, from definition (\ref{eq:defbarrho}) and Eq.\
(\ref{eq:rho-0}) it follows
\ba
\bar{\rho}^{(0)}_n(x)=\psi_n(x) \, .
\la{eq:rho0-1}
\ea

\subsection{Dispersion relation for $F^{\gamma^{*}\gamma\pi}$}
\label{subsec:disper-rel}
The spectral density, discussed above, allows us now to construct
the phenomenological spectral density
\ba \label{eq:PhenDensity}
&&  \rho^{\rm phen}(Q^2,s)
=
   \rho^{\rm h}(Q^2,s)
  +\theta(s-s_0)N_{\rm T}~\rho(Q^2,s)\ ,
\ea
which consists of two parts.
The first term, $\rho^{\rm h}$, (with ``h'' denoting hadronic)
encodes the hadronic content of the spectral density, reexpressed in
terms of the $\gamma^* \rho \to \pi$ transition form factor
$F^{\gamma^*\rho \pi}$:
\ba
 \rho^{\rm h}(Q^2,s)
&=&
 \sqrt{2}f_\rho F^{\gamma^*\rho \pi}(Q^2)\cdot \delta(s-m^2_{\rho})\ ,
\ea
where we assumed that the $\rho$ and $\omega$ resonances have the same
mass and can be represented by a $\delta$-function.
Later on, we are going to show how this simple ansatz
can be improved to include a finite width \cite{Kho99}.
The second term in Eq.\ (\ref{eq:PhenDensity}) represents its
perturbative part
\ba
 \rho(Q^2,s)&=& \frac{\mathbf{Im}}{\pi}
     \left[\,F_\text{QCD}^{\gamma^*\gamma^*\pi}(Q^2,-s)
     \right]\, ,
\label{eq:modeldensity}
\ea
in which $F_\text{QCD}^{\gamma^*\gamma^*\pi}(Q^2,q^2)$ will be computed
according to Eq.\ (\ref{eq:PTformfactor}) (i.e., in convolution form)
including contributions up to the NNLO$_\beta$.

Then, substituting $\rho^{\rm phen}$ in the dispersion relation for
$F^{\gamma^*\gamma \pi}(Q^2,q^2)$ in (\ref{eq:Dispersion})
\cite{Kho99}, we obtain
\begin{equation}
  F^{\gamma^{*}\gamma\pi}(Q^2,q^2)
=
  \frac{\sqrt{2}f_\rho F^{\gamma^*\rho \pi}(Q^2)}{m_{\rho}^2+q^2}
 + N_{\rm T} \int\limits_{s_0}^\infty ds~ \frac{\rho(Q^2,s)}{s+q^2}\, ,
\label{eq:initialD}
\end{equation}
where we have assumed that the hadronic spectral density,
$\rho^{\rm h}$, in the dispersion integral can be approximated by the
expression
\begin{equation}
  \frac{\sqrt{2} f_{\pi}~F^{\gamma^*\rho \pi}(Q^2) }{m_{\rho}^2+q^2}
=
  \int\limits_0^{s_0}\!\frac{\rho(Q^2,s)}{s+q^2}\, ds \, ,
\label{eq:grhopi}
\end{equation}
which determines the structure of $F^{\gamma^*\rho \pi}(Q^2)$.
Here, and below, $s_0=1.5$~GeV${}^2$ denotes the duality interval
in the $\rho$-meson channel.

In order to derive the LCSR for $F^{\gamma^*\rho \pi}(Q^2,M^2)$
\cite{Kho99}, we first perform a Borel transformation of Eq.\
(\ref{eq:grhopi}) with respect to the virtuality $q^2$ and then
insert the result into Eq.\ (\ref{eq:initialD}).
Finally, taking the limit $q^2 \to 0$, we arrive at
\begin{equation}
  F_\text{LCSR}^{\gamma^*\gamma\pi}(Q^2)
=
  \int\limits_0^{s_0}\!\!\frac{ds}{m_\rho^2}\,
  \frac{\mathbf{Im}}{\pi}
     \left[\,F_\text{QCD}^{\gamma^*\gamma^*\pi}(Q^2,-s)
     \right]\,
     e^{\left(m_\rho^2-s\right)/M^2}
   + \int\limits_{s_0}^\infty\!\!\frac{ds}{s}\,
     \frac{\mathbf{Im}}{\pi}
     \left[\,F_\text{QCD}^{\gamma^*\gamma^*\pi}(Q^2,-s)
     \right]\, ,
\label{eq:srggpi}
\end{equation}
where $M^2\approx0.7$~GeV$^2$ is the typical value of the Borel
mass parameter and $m_\rho$ is the $\rho$-meson mass.
The first term in Eq.\ (\ref{eq:srggpi}), which is proportional to
$F^{\gamma^*\rho \pi}$, expresses the hadronic content of the
quasi-real photon, whereas the second term encodes the pointlike
subprocesses governed by QCD perturbation theory.
The spectral density, given by Eq.\ (\ref{eq:defrho}), will allow
us to obtain both parts of
$ F_\text{LCSR}^{\gamma^*\gamma\pi}(Q^2)$; viz.,
\ba
&&  F_\text{LCSR}^{\gamma^*\gamma\pi}(Q^2)
 =  N_T \left\{\frac{1}{m^2_\rho} V(Q^2, M^2)
  + \frac{1}{Q^2} H(Q^2) \right\}\, ,
\label{eq:srggpi2}
\ea
where
\ba
&& V(Q^2, M^2)
=
   \int\limits_0^{s_0}\!\! \rho(Q^2,s)
   e^{\left(m_\rho^2-s\right)/M^2} ds \, ,
\label{eq:Vrho}
\ea
and where
\ba
~~~H(Q^2)
=
   Q^2 \int\limits_{s_0}^\infty\!\! \rho(Q^2,s)\frac{ds}{s}
\label{eq:Hrho}
\ea
in correspondence with our remarks below Eq.\ (\ref{eq:srggpi}).
Employing this notation,
$F^{\gamma^*\rho \pi}(Q^2)$ from Eq.(\ref{eq:grhopi})
can be recast in the form
\ba
\label{eq:Fgammarhopi}
  F^{\gamma^*\rho \pi}(Q^2)
=
  \frac{f_\pi}{3 f_{\rho}} V(Q^2, M^2) \, .
\ea
For the sake of completeness and for future use, we also display the
spectral images of Eqs.\ (\ref{eq:srggpi2}), (\ref{eq:Vrho}),
and (\ref{eq:Hrho}) in terms of the normalized spectral density,
defined in Eq.\ (\ref{def:rhobar}):
\ba
\Ds
    V(Q^2, M^2)
&=&
    \int\limits_{x_0}^1\!\! \Ds \exp\left( \frac{m_\rho^2}{M^2}
  - \frac{Q^2}{M^2}\frac{\bar{x}}{x}\right)
    \bar{\rho}(x,Q^2)\, \frac{dx}{x} \, ,
\label{eq:Vrho-x} \\
H(Q^2)
&=&
   \int\limits_{0}^{x_0}\!\! \bar{\rho}(x,Q^2)\, \frac{dx}{\bar{x}},
\label{eq:Hrho-x}
\ea
where $\Ds x_0=Q^2/(Q^2+s_0)$ and $\Ds s=Q^2\bar{x}/{x}$.
The spectral density $\bar{\rho}_n$ beyond the LO will be
constructed and discussed in the next section.
Let us remark at this point that in the following we are going to
consider also mixed forms of these sets of variables, i.e.,
a spectral density which depends on both variables $s$ and $x$.
This should not cause any confusion because it is understood that
one has to replace each time the appropriate variable, i.e.,
either $ s=Q^2\bar{x}/{x}$ or $ x =Q^2/(Q^2+s)$.

\section{Structure of the spectral density beyond LO}
\label{sec:struc-spec-dens}
This section extends our calculation of the spectral density beyond the
leading order of perturbation theory.
In the following two subsections we will take into account the
radiative corrections at the NLO and also at the NNLO; in the latter
case only the $\beta_0$-proportional terms will be included.

\subsection{Spectral density in NLO}
\label{subsec:spec-dens-NLO}
Let us first write down the final result for $\bar{\rho}^{(1)}_n$ for
any index $n$ and then proceed with a systematic discussion of its
structure:
\begin{subequations}
\label{eq:rho1}
\begin{eqnarray}
  \frac{1}{C_{\rm F}}\bar{\rho}^{(1)}_n\left(x;s/\mu^2_{\rm F}\right)
 = &&
  \!\!\!\!\!  \left\{-3\left[1+ v^{b}(n)\right]
         +\frac{\pi^2}{3}-\ln^2\left(\frac{\bar{x}}{x}\right)
         +2v(n) \ln\left(\frac{s}{\mu^2_{\rm F}}\right)
 \right\} \psi_n(x)
\la{eq:rho1a}\\
&& -2v(n) \int_0^{\bm{\bar{x}}} du
  \left[\frac{\psi_n(u)-\psi_n(\bm{\bar{x}})}{u-\bm{\bar{x}}}\right]
\la{eq:rho1c}\\
&& -2\int_x^1 du \left[\frac{\psi_n(u)-\psi_n(x)}{u-x}\right]
  \ln\left(1-\frac{x}{u}\right)
 +(x \to \bar{x})
\la{eq:rho1b}\, .
\end{eqnarray}
\end{subequations}
The partial cases $n=0, 2, 4$ coincide, after some algebraic
manipulations, with the results obtained before by Schmedding
and Yakovlev (SY) \cite{SchmYa99}.\footnote{The corresponding
explicit expressions, denoted $A_0, A_2$, and $A_4$, can be found
there.}
The above expression---besides being valid for any $n$---also reveals
how the radiative corrections manifest themselves in its various terms,
as we will now explain.

The quantities $~v^{b}(n)$ and $v(n)$ in (\ref{eq:rho1a}) and
(\ref{eq:rho1c}) are the eigenvalues of the corresponding parts of the
one-loop ERBL evolution kernel $V^{(0)}$, notably, $V^{b}_+$,
and $~\left(V^{b}_+ + V^{a}_+\right)$.
These are given by
\ba
  V^{(0)}(x,y)
&=&
  C_{\rm F}  2\left[ V_+^{a}(x,y) + V_+^{b}(x,y)\right]
=
  C_{\rm F} V_+(x,y) \, ,\\
  V^{(0)}(x,y)\otimes\psi_n(y)&=&C_{\rm F} 2v(n)\psi_n(x)\, ,
\la{eq:V}
\ea
where the eigenvalues of the partial kernels $V^{a,b}$ are obtained
from
\begin{subequations}
\label{eq:V0}
\ba
  V_+^a(x,y) \otimes\psi_n(y)&=&v^a(n)\psi_n(x),~v^a(n)
=
  \frac{1}{(n +1)( n+2)} -
  \frac{1}{2}\, ,
\la{eq:va}\\
  V_+^b(x,y) \otimes\psi_n(y)&=&v^b(n)\psi_n(x),
  ~v^b(n)
=
  2 \left[\psi(2)- \psi(2+n)\right]
\la{eq:vb}\, .
\ea
 \end{subequations}
Here $\psi(z)= d\ln\Gamma(z)/dz$ and
$V_+(x,y)=V(x,y) -\delta(x-\bm{y})\int^1_0 V(t,\bm{y})~dt$.
The latter definition reflects the vector-current conservation,
while further details pertaining to the above equations are relegated
to Appendix \ref{app:Vstructure}.
Note that the term (\ref{eq:rho1c}) originates from the logarithm in
$T_{1}$ [cf.\ Eq.\ (\ref{eq:L-n})].
Therefore, its contribution to the discontinuity contains, in
comparison with $\rho^{(0)}$ in (\ref{eq:rho-0}), a new element
which is discussed in Appendix~\ref{app:Tstructure}, item 2,
Eq.\ (\ref{A12}).
The corresponding kernel in Eq.\ (\ref{eq:rho1c}) maps each monomial
term again onto a monomial one, e.g.,
\be \label{4.3}
\int_0^x du \left[\frac{u^n - x^n}{u-x}\right]
 \equiv\left[\frac{\theta(x>u)}{u-x}\right]_{+(x)}\otimes u^n
=\left[\psi(n+1)-\psi(1)\right]x^n \ ,
\ee
where a notation deviating from the usual $+$ prescription
\ba
  \left[f(x,u)\right]_{+(x)} = f(x,u) - \delta(x-u)\int_0^1 f(x,t)\ dt
\label{eq:def+y}
\ea
has been used.
Keeping in mind that, ultimately, we have to integrate the spectral
density $\rho(x)$ over $x$, it is particularly useful to recast Eq.\
(\ref{eq:rho1c}) in terms of an expansion over an orthogonal
polynomial basis, e.g., over the eigenfunctions $\{\psi_n\}$:
\ba \label{4.b}
   \int_0^{\bm{\bar{x}}} du
  \left[\frac{\psi_n(u)-\psi_n(\bm{\bar{x}})}{u-\bm{\bar{x}}}\right]
=
  \sum^n_{l=0,1,\ldots} b_{n l}~\psi_{l}(x) -3\bm{\bar{x}} \, .
\ea
The expansion coefficients $b_{n l}$ are given in Appendix A, item 2.

As regards the term (\ref{eq:rho1b}), it originates from the kernel
$g(y,u)$, introduced in \cite{Mul94} and discussed in \cite{MMP02}
(with some technical details being provided for the reader's
convenience in Appendix \ref{app:Tstructure}, item 2).
This kernel, termed $g$ in \cite{Mul94}, is not diagonal with respect
to the $\{\psi_n\}$-basis and is responsible for the apparent breaking
of the conformal symmetry in the $\overline{\rm MS}$-scheme
\cite{Mul94}.
Notice that also the kernel in Eq.\ (\ref{eq:rho1b}) maps each term
$\psi_n$ onto a sum of $\psi_l(y),~l\leq n $.
Therefore, (\ref{eq:rho1b}) can be cast in the form of the following
algebraic expansion
\ba
\label{4.G}
  \int_x^1 \left[\frac{\psi_n(u)-\psi_n(x)}{u-x}\right]
  \ln\left(1-\frac{x}{u}\right) du + (x \to \bar{x})
= \sum^n_{l=0,2,\ldots}~G_{nl}~\psi_l(x) \ .
\ea
Note that the various terms in the spectral density $\bar{\rho}^{(1)}$
have a one-to-one correspondence to the kernel $V^{(0)}$ and its
elements, and partly also to the kernel $V^{(1)}$ via its element
$g$.

(i) For the special case $n=0$, $\psi_0(x)=6 x\bar{x}$ so that
the dependence of $\bar{\rho}^{(1)}_0$ on the factorization scale
$\mu^2_{\rm F}$ disappears, owing to the fact that the
asymptotic DA does not evolve in this approximation.
Indeed, in this case, the following chain of evident simplifications
is induced
\ba
&&v^b(0)=v^a(0)=v(0)=0;
\la{eq:rho1-0a} \\
&&\left[
        \frac{\theta(u>x)}{u-x}\ln\left(1-\frac{x}{u}\right)
  \right]_{+(x)}
\otimes
  \psi_0(u)
  +(x \to \bar{x}) = \psi_0(x)\, .
\la{eq:rho1-0b}
\ea
Substituting Eqs.\ (\ref{eq:rho1-0a}), (\ref{eq:rho1-0b}) in
expressions (\ref{eq:rho1a})--(\ref{eq:rho1b}), one arrives at
\ba
  \bar{\rho}^{(1)}_0(x)
=
   C_{\rm F} \left[
        -3-2+\frac{\pi^2}{3}-\ln^2\left(\frac{\bar{x}}{x}\right)
  \right]\psi_0(x)\, .
\label{eq:bar-rho}
\ea
This expression agrees with the result obtained in
\cite{SchmYa99} for $\rho^{(1)}_0(Q^2,s)$.

(ii) For the general case of an arbitrary $n$, the
``nondiagonal'' (in $\psi_n$) part of $\bar{\rho}^{(1)}$ in the
second line of the expression (note that $s=Q^2 \bar{x}/x$)
\ba
  \frac{1}{C_{\rm F}} \bar{\rho}^{(1)}_n\left(x;\bm{Q^2}/\mu^2_{\rm F}\right)
&=&
            \left\{
                  -3\left[1 + \bm{v^{b}(n)}\right]+\frac{\pi^2}{3}
                  -\ln^2\left(\frac{\bar{x}}{x}\right)
                  +2v(n)\ln\left(\frac{\bar{x}}{x} \right)
                  +2v(n) \ln\left(\frac{Q^2}{\mu^2_{\rm F}}\right)
           \right\}
\nonumber\\
&& \times         \psi_n(x)
        -  2\left[
                  \sum^n_{l=0,2,\ldots}~G_{nl}~\psi_l(x)
                  +v(n)\left( \sum^n_{l=0,1,\ldots} b_{n l}\psi_l(x) \bm{-3\bar{x}}\right)
           \right]
\la{eq:rho1-n}
\ea
has been rewritten in terms of the known coefficients
$G_{nl}$ and $b_{nl}$, supplied in Appendix \ref{app:Tstructure}
in terms of Eqs.\ (\ref{A11}) and (\ref{A17}).
This way, we have achieved that Eq.\ (\ref{eq:rho1-n}) is a purely
algebraic expression for $\rho$.
To return to the $n=0$ case, one should set in Eq.\ (\ref{eq:rho1-n})
$v^a(0)=v(0)=0$ and $G_{00}=1$.
Expressions (\ref{eq:rho1a})--(\ref{eq:rho1b}), or, equivalently,
(\ref{eq:rho1-n}), provide an effective tool for analyzing any model
of the pion DA within the LCSR approach.
At this point it is worth comparing the NLO contribution, originating
from two different approaches, with respect to the case when
one of the photons becomes real.
First, we have the expression obtained directly from
the factorization formula in Eq.\ (\ref{Str-T1-log}) for $q^2 =0$
\cite{KMR86,MuR97,BMS02}, i.e.,
\ba \la{psi0-fact}
  Q^2 T_1(Q^2, 0,x)\otimes\psi_0(x)
=
  C_{\rm F} \int^1_0
  \left[\ln^2(x)- \frac{x}{1-x}\ln(x)-9 \right]\frac{\psi_0(x)}{x}dx
=
  -15 C_{\rm F}\ .
\ea
Second, starting from the dispersion relation given by
Eq.\ (\ref{eq:Hrho-x}) for $H$, one can return to the previous
expression by setting $s_0=0,~x_0=1$ to get
\ba \la{psi0-disp}
  \int^1_0 \frac{\bar{\rho}^{(1)}_0(\bar{x})}{x} d x
=
  C_{\rm F} \int^1_0
  \left[\frac{\pi^2}{3}-\ln^2\left(\frac{\bar{x}}{x}\right)
  -5 \right]\frac{\psi_0(x)}{x}
=
  -15 C_{\rm F} \ ,
\ea
where we have used $\bar{\rho}^{(1)}_0$ from Eq.\ (\ref{eq:bar-rho}).
The outcome of both expressions is the same, though the integrands are
different.
However, the $\ln^2$--terms in both formulas (\ref{psi0-fact}) and
(\ref{psi0-disp}) have the same origin: notably, the $g$-kernel
contributing to $T_1$ [cf.\ (\ref{Str-T1})--(\ref{g})],
as one appreciates from the relation
$$
 (g_+ \otimes \psi_n)(x)
=
 \left[\frac{\pi^2}{3}-\ln^2\left(\bar{y}/y\right)\right]
 \psi_n(x) + \mbox{less singular terms in } x .
$$
The crucial observation here is that the dispersion method yields an
expression that contains the leading squared logarithm with a negative
sign---in contrast to the result one finds with the factorization
approach [cf.\ (\ref{psi0-fact})].
In this second case, it is more involved to show \cite{MuR97} that the
leading logarithm provides suppression in the relevant integration
region and can therefore be associated with Sudakov effects.

\subsection{Spectral density in NNLO.
            $\beta_0$-proportional contributions}
\label{subsec:spec-dens-NNLO}
The $\beta$-dependent part of the partial spectral density,
$\bar{\rho}^{(2 \beta)}_{n}$,
can be obtained from the corresponding part of the whole amplitude
$T_{2}$, given by Eqs.\ (\ref{T2beta-n}), i.e., from
$T_{\beta}\otimes \psi_n$.
The calculation of the discontinuity of the latter expression is a
rather technical task and is, therefore, relegated to the two
Appendices \ref{app:Tstructure} and \ref{App-NNLOelements}.
It is important to realize that the structure of the spectral density,
$\bar{\rho}^{(2 \beta)}_{n}$, resembles the structure of the
analogous term in the NNLO$_\beta$-amplitude that is proportional to
 $T_{\beta}\otimes \psi_n$, as one may appreciate by comparing the
 following two expressions:
\begin{subequations}
\label{eq:T2-n}
\ba
  T_{2}\otimes \psi_n
&\to&
  b_0 \cdot T_{\beta}\otimes\psi_n(q^2,Q^2,\mu_{\rm F}^2,\mu^2_{\rm R})
=
  b_0 \left[
            \ln\left(\frac{\mu_{\rm R}^2}{\mu_{\rm F}^2}\right)
            T_{1}\otimes \psi_n~~~~~~~~~~~~~~~~~~~~~~~~~~~
\right.
\label{eq:T1inT2-n}\\
&& \left. +~C_{\rm F} T_{0}\otimes
     \left\{ {\cal T}^{(2)}_{\beta}
             +{\rm Ln}\! \cdot \! \left[(2V^{(1)}_{\beta})_+
             -{\cal T}^{(1)}\right]
             - \frac{1}{2}{\rm Ln}^2 \! \cdot \!
    (2 v(n)) \right\}
   \otimes \psi_n \right] ,
\label{eq:T2inT2-n}
\ea
 \end{subequations}
\begin{subequations}
 \label{eq:rho2-n}
  \ba
   \bar{\rho}^{(2)}_n
&\to&
   b_0 \cdot \bar{\rho}^{(2 \beta)}_{n}
   \left(x;s,\mu_{\rm F}^2,\mu^2_{\rm R}\right)
=
   b_0  C_{\rm F}
   \left[\ln\left(\frac{\mu_{\rm R}^2}{\mu_{\rm F}^2}\right)
         \bar{\rho}^{(1)}_n(x;s/\mu_{\rm F}^2)~~~~~~~~~~~~~~~~~~\right.
\label{eq:rho1inrho2-n} \\
&& \left.
\phantom{
         b_0 \cdot \bar{\rho}^{(2\beta)}_{n}
         \left(x;s/\mu_{\rm F}^2,\mu_{\rm R}^2 \right)
         \ln\left(\frac{\mu_{\rm R}^2}{\mu_{\rm F}^2}\right)
        }
   ~~~~~~~~~~+ \bar{R}^{(2)}_n(x;s/\mu^2_{\rm F})
   \right]\, .
\label{eq:Rho2inrho2-n}
\ea
\end{subequations}
For convenience, we have made use of the abbreviation
${\rm Ln}$ [cf.\ (\ref{eq:L-n})],
omitting arguments and separating for emphasis such terms
from other functions by a dot.
Recalling the results of the previous subsection, one appreciates that
the spectral density $\bar{\rho}^{(1)}_n(x,s/\mu^2_{\rm F})$ follows
from the first term on the RHS of Eq.\ (\ref{eq:T1inT2-n}).
On the other hand, the new contribution
$\bar{R}^{(2)}_n(x,s/\mu^2_{\rm F})$
in Eq.\ (\ref{eq:Rho2inrho2-n}) derives from Eq.\ (\ref{eq:T2inT2-n})
and represents one of the main results of this investigation.

To continue, consider the important partial case $n=0$ for which the
corresponding expression for $\bar{\rho}^{(2 \beta)}_{0}$ becomes
significantly simplified.
Indeed, the leading logarithmic term contributing to $\bar{R}^{(2)}_0$
and stemming from ${\rm Ln}^2(y)$ [cf.\ (\ref{eq:T2inT2-n})] cancels
out because it is proportional to $v(n)$.
Actually, this is a general property of the leading logarithmic terms
in all orders of the expansion that first reveals itself at the NLO
level---see Eq.\ (\ref{eq:rho1a}).
On the other hand, the subleading logarithmic term
$\sim \ln\left(s/\mu^2_{\rm F}\right)$
survives, because it originates from the $V^{(1)}_{\beta}$--element of
the $V^{(1)}$ kernel and from the ${\cal T}^{(1)}$--NLO element of the
hard-scattering amplitude that does not vanish for the $\psi_0$
harmonic [recall the discussion after Eq.\ (\ref{eq:T-beta})].
Thus, the final result for $\bar{R}^{(2)}_0$ reads
\ba
\bar{R}^{(2)}_0(x;s/\mu^2_{\rm F})
=
  \left({\cal T}^{(2)}_{\beta}\otimes \psi_0\right)(x)
        + C_2(x)
        +\ln\left(\frac{s}{\mu^2_{\rm F}}\right)\cdot C_1(x)\, ,
\label{eq:R2}
\ea
where the individual ingredients of this equation are the following
\ba
  C_1(x)
&=&
  \left[(\bm{V^{(1)}_{\beta}})_+ - {\cal T}^{(1)}_{\rm F}\right]
  \otimes \psi_0
=
  -6\left(\bar{x} \ln(\bar{x}) + x \ln(x)\right) \, ,
\label{eq:C1}  \\
  C_2(x)
&=&
  - \int_0^{\bm{\bar{x}}} du \left[\frac{C_1(u)-C_1(\bm{\bar{x}})}{u-\bm{\bar{x}}}\right]
  \nonumber\\
&=&
 \bm{-3x} \ln^2(x)+\pi^2\bar{x}+
\bm{ 6\left(\bar{x}\ln(\bar{x})+x\ln(x)-x \text{Li}_2(\bar{x})\right)},
\label{eq:C2}
\ea
\ba
  {\cal T}^{(2)}_{\beta}\otimes \psi_0
&=&
   x\bar{x}
   \Big\{
          30\left[\text{Li}_3(x)+\text{Li}_3(\bar{x})\right]
         -6\left[ \ln(\bar{x}){\rm Li_2}(\bar{x})+\ln(x){\rm Li_2}(x)
           \right]
         -\left[\ln^3(\bar{x})+\ln^3(x)\right]
\nonumber \\
&&
  -5\ln^2\left(\frac{\bar{x}}{x}\right)
  +\left[\ln(\bar{x})+\ln(x)\right]
   \left(3\ln(\bar{x})\ln(x)-2\pi^2\right)
  -72\zeta(3)+\frac{5}{3}\pi^2-7
   \Big\}
\nonumber \\
&& +\frac{19}{2}\left[\bar{x}\ln(\bar{x})+x\ln(x)\right]
   +\frac{3}{2}\left[\bar{x}\ln^2(\bar{x})+x\ln^2(x)\right]\, .
\label{eq:Tbeta}
\ea
\begin{figure}[ht]
\centerline{\includegraphics[width=0.6\textwidth]{
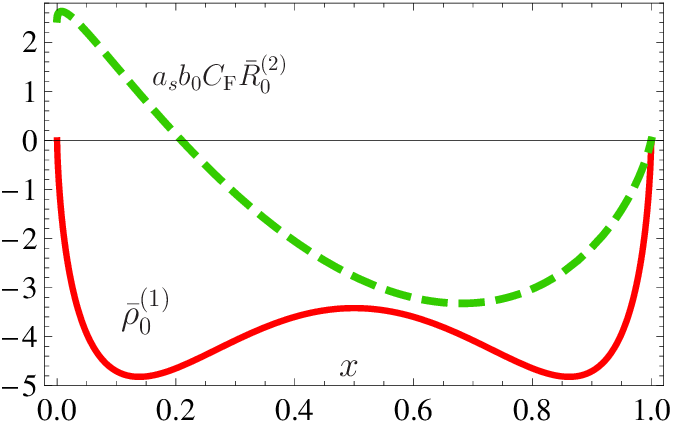}}
\vspace*{-3mm}
\caption{\footnotesize The dashed green line shows the expression
          $a_s(\mu^2_{\rm F})~\bar{\rho}^{(2)}_0
          =
          a_s(\mu^2_{\rm F})~b_0 C_{\rm F}\bar{R}^{(2)}_0(x,\bar{x}/x)$
          at the typical CLEO reference scale (labeled by the acronyms
          of Schmedding and Yakovlev \cite{SchmYa99})
          $Q^2=\mu^2_{\rm F}=\mu^2_{\rm SY}=(2.4~\text{GeV})^2$,
          whereas the solid red line represents
          $\bar{\rho}^{(1)}_0(x)$ in (\ref{eq:bar-rho}).}
\label{F:Fig3}
\end{figure}
To derive Eq.\ (\ref{eq:R2}), we have made use of
Eqs.\ (\ref{rho0}), (\ref{A12}) for the spectral density elements and
also (\ref{eq:T2inT2-n}) in conjunction with
Eqs.\ (\ref{C11}) and (\ref{C18}).
The coefficient $C_1$ in front of
$\ln\left(s/\mu^2_{\rm F}\right)$ in Eq.\ (\ref{eq:R2})
accumulates those contributions responsible for the breaking of the
conformal symmetry in the $\overline{\rm MS}$-scheme owing to
$\dot{V}$ \cite{MR85} and $g$ \cite{Mul94}.

Note that just this term leads to the breaking of the
$(x\leftrightarrow \bar{x})$-symmetry of the spectral density
$
 \Ds \bar{R}^{(2)}_0\left(x,\frac{Q^2}{\mu^2_{\rm F}}\frac{\bar{x}}{x}
                    \right)
$
[recalling that $\Ds s= Q^2 \bar{x}/x$].
In Fig.\ \ref{F:Fig3}, we compare the contributions to the spectral
density from the NLO and NNLO$_\beta$ at one single scale
$Q^2=\mu^2_{\rm F}=\mu^2_{\rm SY}$,
the latter scale $\mu^2_{\rm SY}=(2.4 ~\gev{} )$$^{2}$ corresponding to
the typical average momentum \cite{SchmYa99} measured by the CLEO
Collaboration \cite{CLEO98} in the $Q^2$ region
$\left[1.5 - 8\right]$~GeV$^{2}$.
This is done in terms of
$a_s(\mu^2_{\rm F}) b_0 C_{\rm F}\bar{R}^{(2)}_0(x, \bar{x}/x)$ for the
NNLO$_\beta$ spectral density (dashed green line) in comparison
with $\bar{\rho}^{(1)}_0(x)$ calculated at the NLO (solid red line).

To render our presentation more transparent, we have relegated the
discussion of the elements
${\cal T}^{(2)}_{\beta}, V^{(1)}_{\beta}, {\cal T}^{(1)}$,
contributing to the amplitude $T_{2}$
[see Eqs.\ (\ref{eq:R2}), \ (\ref{eq:C1})],
to the Appendices \ref{app:Tstructure} and
\ref{app:Vstructure}.
The expressions for the partial densities
$\bar{R}^{(2)}_n(x,s/\mu^2_{\rm F})$,
appearing as convolutions of these elements with $\psi_n$, are
supplied in Appendix \ref{App-NNLOelements}.

\section{Size of the NNLO contribution to the form factors}
\label{sec:FF-results}
\begin{figure}[t]
 \centerline{\includegraphics[width=0.45\textwidth]{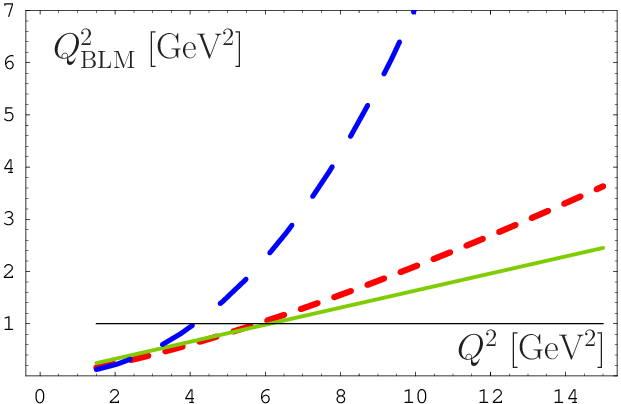}~~~~~~
             \includegraphics[width=0.48\textwidth]{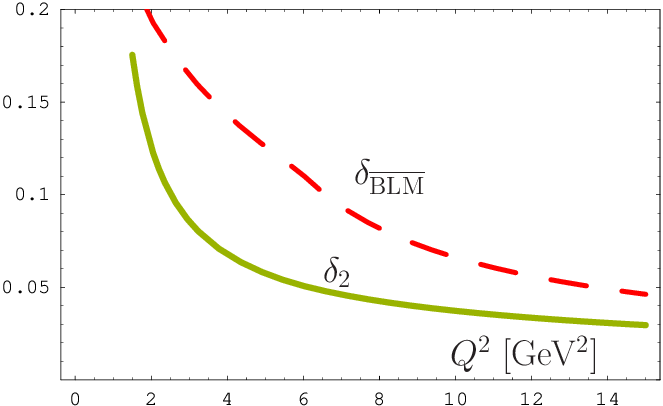}}
 \vspace*{-1mm}
 \caption{\footnotesize Left:~The dashed (red) line in the middle
 represents $Q_{\rm BLM}^2(Q^2)$,
 whereas the solid (green) line corresponds to the result
 $\tilde{Q}^2_{\rm BLM}=Q^2 \exp(-1.811)$ in \cite{MMP02},
 and the upper dashed (blue) line marks the BLM scale for the
 $V$-part [cf.\ Eq.\ (\ref{eq:srggpi2})]
 of the form factor describing the $\gamma^* \rho \pi$ transition.
 Right: The upper dashed (red) line represents the BLM ratio
 $\delta_{\overline{\rm BLM}}$, while the lower (green) line denotes
 $\delta_2$ in Eq.\ (\ref{eq:NNLOratio}).
 }
 \label{F:Fig4}
\end{figure}

In this section we discuss the effects of the NNLO$_\beta$ corrections
to the transition form factors for the $\rho$ and $\pi$ mesons,
relying upon the BLM prescription and its modifications.
Employing the results obtained in the previous section,
we are going to simplify the expressions for the spectral density
by adopting the so-called `default' scale setting
$\mu^2_{\rm F}=\mu_{\rm R}^2=Q^2$.
Then, the expression for $\rho^{(2 \beta)}_{n}$, given by Eq.\
(\ref{eq:Rho2inrho2-n}), reduces to
\ba
  \rho^{(2 \beta)}_{n}(x;s,Q^2,Q^2)
=
  C_{\rm F} \bar{R}^{(2)}_n(x,\bar{x}/x)
\ea
(see Fig.\ \ref{F:Fig3}).
Inserting this result in
Eqs.\ (\ref{eq:Vrho-x})--(\ref{eq:Hrho-x}) and performing the
integration over $x$, one finally finds for their sum
[Eq.\ (\ref{eq:srggpi2})] an expression that contains
$F_{\beta}(Q^2)$ at the NNLO level.
The final result for the whole form factor reads
\ba \label{eq:generNNLO}
&&  F_{\rm LCSR}^{\gamma^*\gamma\pi}(Q^2)
=
    F_0(Q^2) + a_s(Q^2)~F_1(Q^2)
    + \left(a_s(Q^2)\right)^{2} b_0 F_{\beta}(Q^2) \, .
\ea
It turns out that the NNLO$_\beta$ contribution, calculated here,
is negative (dashed line in Fig.\ 3).
Hence, taken together with the already known NLO contribution,
which is also negative, the total effect of the radiative
corrections at the considered level of the perturbative expansion
is to decrease the magnitude of the form factor.
Following the BLM procedure, the last term in Eq.\ (\ref{eq:generNNLO})
determines the ``shift'' of the scale in the argument of the running
coupling from the value $Q^2$ to the BLM-scale \cite{BLM83},
$Q^2_{\rm BLM}$, according to
\ba
&&
  Q^2_{\rm BLM}(Q^2)
=
    Q^2 \exp\left\{- \frac{F_{\beta}(Q^2)}{F_1(Q^2)}\right\}\, .
    \label{Eq:BLMscale}
\ea
As a result,
$a_s(Q^2) \to a_s(Q^2_\text{BLM})> a_s(Q^2)$
at $Q^2_{\rm BLM}(Q^2) < Q^2 $
and, hence, the form factor given by Eq.\ (\ref{eq:generNNLO})
assumes the BLM--improved form
\ba \label{eq:BLMformfactor}
  F_{\text{LCSR}}^{\gamma^*\gamma\pi}(Q^2)
\to
  F_{\text{BLM}}^{\gamma^*\gamma\pi}(Q^2)
=
  F_0(Q^2) + a_s(Q_\text{BLM}^2)F_1(Q^2)\, .
\ea
The main contribution to $F_\text{LCSR}^{\gamma^*\gamma\pi}$ is
provided by the asymptotic DA, $\psi_0$.
The associated BLM scale is
$Q^2_{\rm BLM}(Q_{\rm min}^2)\equiv Q^{2}_{\star}\approx 1$~\gev{2}
(see Fig.\ \ref{F:Fig4}, left panel).
This scale may be considered as the borderline for applying
perturbative QCD, defining this way some minimal scale for the BLM
scheme---denoted $Q_{\rm min}^2$.
It is remarkable (though accidental) that, as mentioned above, this
scale corresponds approximately to
$\mu_{\rm SY}^2=5.76$~\gev{2}.
Obviously, below this particular scale, the BLM prescription,
expressed through Eq.\ (\ref{Eq:BLMscale}), would entail a
renormalization scale that will be out of the region where
perturbation theory can be safely applied.

\vspace{-1mm}

The scale $Q^{2}_{\star}$, determined above, still belongs to
the perturbative regime and, therefore, the ratio
\be
\label{eq:BLMratio}
 \delta_\text{BLM}(\mu_{\rm SY}^2)\equiv
 \left[F_\text{LCSR}^{\gamma^*\gamma\pi}(\mu_{\rm SY}^2)
 -F^{\gamma^*\gamma\pi}_{\text{BLM}}(\mu_{\rm SY}^2)\right]/
F_\text{LCSR}^{\gamma^*\pi}(\mu_{\rm SY}^2)\approx 0.11
\ee
can serve as a crude measure for the relative weight of the NNLO
contribution.
In Fig.\ \ref{F:Fig4} (left panel), we show
$Q^2_{\rm BLM}(Q^2)$ vs.\ $Q^2$
(dashed red line in the middle), following from Eq.\
(\ref{Eq:BLMscale}), in comparison with a simpler linear dependence
$\tilde{Q}^2_{\rm BLM}=Q^2 \exp(-1.811)$ (solid green line)
that results from the standard factorization formula of the
perturbative approach employed in \cite{MMP02}.
One observes that both results are rather close to each other in the
important CLEO-data region.
To be able to apply the BLM procedure below the minimal
scale $Q^2_{\rm min}$, we shall use a somewhat improved version
of this procedure---termed $\overline{\rm BLM}$---introduced in
\cite{BPSS04}.
Above the perturbation-theory borderline $Q^2\geq Q_{\rm min}^2$,
this modified BLM procedure coincides on the RHS of
Eq.\ (\ref{eq:BLMformfactor}) with the standard one.
But for $Q^2< Q_{\rm min}^2$, the BLM scale is frozen at $Q^2_{\star}$
and the expanded expression contains only a tail of the NNLO$_\beta$
correction provided by the third term in the equation below
\be
  F_{\overline{\rm BLM}}^{\gamma^*\gamma\pi}(Q^2)
=
    F_0(Q^2)
  + a_s(Q_{\star})F_1(Q^2)
  + a_s^2(Q^2_{\star})b_0 \left( F_{\beta}(Q^2)
  - F_1(Q^2)\ln(Q^2/Q^2_{\star})\right)\ .
\ee
Substituting this expression for
$F^{\gamma^*\gamma\pi}_{\text{BLM}}$
into Eq.\ (\ref{eq:BLMratio}), one obtains the quantity
$\delta_{\overline{\rm BLM}}$, shown in the right panel of Fig.\
\ref{F:Fig4}.
The numerical value, estimated in Eq.\ (\ref{eq:BLMratio}), turns out
to overestimate the size of the NNLO contribution, as one observes
from this figure.
In mathematical terms this becomes evident by glancing at the ratio
\be \label{eq:NNLOratio}
 \delta_{2}(Q^2)\equiv
 |a_s^{2}(Q^2) b_0 F_{\beta}(Q^2)|/
F_\text{LCSR}^{\gamma^*\gamma\pi}(Q^2)
\ee
and recalling that the magnitude of this contribution to the total
form factor is a few times smaller than
$\delta_{\overline{\rm BLM}}$
at the moderate values of $Q^2$ characterizing the CLEO-data region.
The size of the NNLO correction seems to be rather important
and at the level of about 10\% at low $Q^2$,
even by taking it into account only in the incomplete form of
Eq.\ (\ref{eq:generNNLO}).
On the other hand, at higher momenta, inspection of the $\delta_{2}$
behavior in the right panel of Fig. \ref{F:Fig4} reveals that the
size of the corrections rapidly decreases to the level of 5\% around
the scale $\mu_{\rm SY}^2$.

\section{Predictions and comparison with experimental data}
\label{sec:data-comp}
This section contains a discussion of the implications of our
theoretical findings on the transition form factors
$F^{\gamma^{*}\gamma\pi}$ and $F^{\gamma^{*}\rho\pi}$ vis-a-vis
the experimental data for the former.
To understand the influence of the NNLO radiative correction
on these hadronic observables, we have to analyze its relative
weight with respect to the NLO contribution.
Moreover, we have to discuss the interplay between perturbative
corrections and nonperturbative ingredients, notably, the quark
virtuality $\lambda_{q}^2$ and the twist-four scale $\delta^2$.
This discussion can be further substantiated by comparing the
calculated photon-to-pion transition form factor with the
available experimental data from measurements by the CLEO
\cite{CLEO98} and the CELLO \cite{CELLO91} Collaborations.
From this comparison, we can extract valuable information as to what
extent our calculation can describe the data in the whole measured
$Q^2$ region.
From the theoretical point of view, we can use these data in order
to estimate what is still missing on the theoretical side.
We will focus below not on the exact phenomenological description of
the mentioned data, but analyze instead the ramifications they
imply on the theoretical approach and its various elements.

\subsection{Nonperturbative input}
\label{subsec:nonperturbative}
We discuss first the nonperturbative ingredients of our analysis.
The main one is the leading twist-two pion DA, $\varphi_{\pi}^{(2)}$,
which can be derived with the help of various methods.
These include---among others---QCD sum rules and lattice simulations.
In addition, one has to model the twist-four component of the pion DA,
see, e.g., the discussion in \cite{BMS03,Ag05b,BMS05lat}.
Lacking a complete derivation of the full pion DA from first principles
of QCD, we are actually forced to reverse-engineer its structure from
calculations of its first few moments $\langle \xi^{n}\rangle_{\pi}$.
To be more precise, one can calculate the moments of
$\varphi_{\pi}^{(2)}$ with standard QCD SR \cite{CZ84} and also with
those employing nonlocal condensates (NLC)
\cite{MR89,MR92,BM95,BM98,BMS01}.
On the other hand, one can use LCSR to analyze the high-precision CLEO
data \cite{CLEO98} on $F^{\gamma^{*}\gamma\pi}$ and extract rather
strict constraints on the first Gegenbauer coefficients $a_2$ and $a_4$
\cite{SchmYa99,BMS02,BMS03}.
More recently, two independent Collaborations have published results for
the first coefficient $a_2$ by measuring the first moment
$\langle \xi^{2}\rangle_{\pi}$ of $\varphi_{\pi}^{(2)}$ on the lattice
\cite{DelDebbio05,Lat06,Lat07}.
The lattice calculation of the second coefficient $a_4$ ( or,
equivalently, the fourth moment
$\langle \xi^{4}\rangle_{\pi}$) is still lacking, but a compatibility
region between the CLEO data \cite{CLEO98} and the $a_2$-lattice
constraints was worked out in \cite{Ste08} to predict a rather
narrow interval for the $\langle \xi^{4}\rangle_{\pi}$ values.

\begin{figure}[t]
 \centerline{\includegraphics[width=0.47\textwidth]{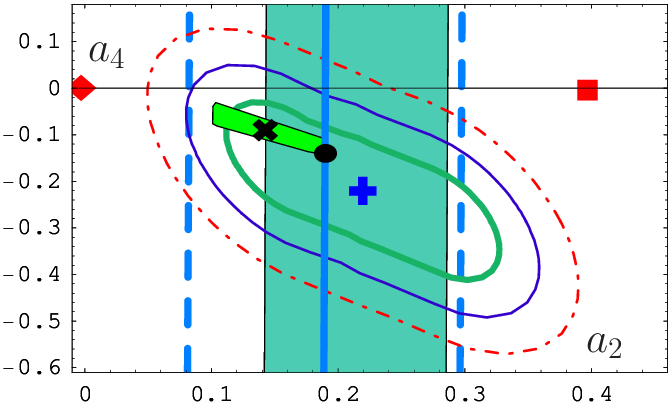}~~~~~~~
             \includegraphics[width=0.47\textwidth]{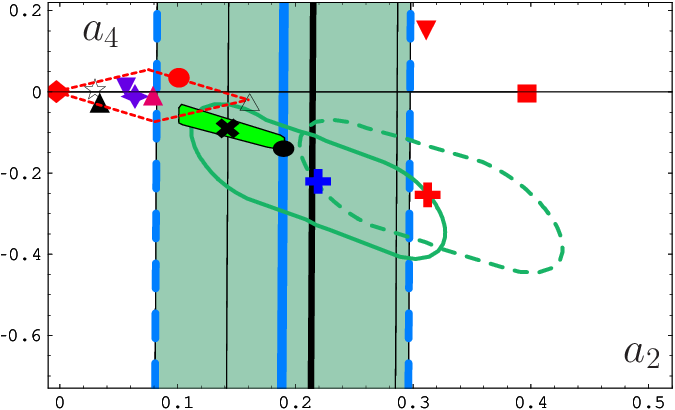}}
 \vspace*{-3mm}
 \caption{\footnotesize Left:
Comparison of the CLEO-data constraints on
$F^{\gamma^{*}\gamma \pi}(Q^2)$ in the ($a_2$, $a_4$) plane
at the scale $\mu_{\rm SY}^2$ in terms of error regions around the
BMS best-fit point {\BluTn{\ding{58}}} \protect{\cite{BMS02,BMS03}},
using the following designations:
$1\sigma$ (thick solid green line);
$2\sigma$ (solid blue line);
$3\sigma$ (dashed-dotted red line).
Two recent lattice simulations, denoted by vertical dashed lines
\protect\cite{Lat06} and solid ones \cite{Lat07} are also shown
together with predictions obtained from nonlocal QCD sum
rules (slanted green rectangle) \protect\cite{BMS01}---the latter
corresponding to the vacuum quark virtuality
$\lambda^2_q=0.4$~GeV$^{2}$.
Right:
CLEO data in comparison with various theoretical models and lattice
results listed in Table \ref{tab:collage}.
The dashed green $1\sigma$ ellipse (together with the shifted best-fit
point \RedTn{\ding{58}}) describes the effect of including the
twist-four contribution to the pion DA via renormalons
\protect\cite{BMS05lat}.
}
\label{F:DAmodels}
\end{figure}

We provide below a short overview of our present knowledge of
$\varphi_{\pi}^{(2)}$ from different sources, omitting specific details
for which we refer the interested reader to the original literature.
The models shown below in the figures are summarized in Table
\ref{tab:collage}.
Using QCD sum rules with NLC, we derived a ``bunch'' of admissible pion
DAs \cite{BMS01}, taking into account in the expansion
(\ref{eq:Gegenbauer}) only the first two terms with $a_2$ and $a_4$
(details can be found in \cite{BMS04kg}).
Pion DA models, like the Chernyak-Zhitnitsky (CZ) model \cite{CZ84}, or
the Braun-Filyanov (BF) \cite{BF89} one, also used these two harmonics
for modeling the pion DA.
Note, however, that in the NLC approach \cite{BMS01} this is not the
result of an arbitrary truncation of the Gegenbauer expansion after
$n=4$, but follows from the fact that all calculated
higher-order coefficients up to $n=10$
turn out to be compatible with zero.
Hence, from a pragmatic point of view, in order to capture the main
characteristics of the pion DA, it is sufficient to restrict the
analysis of the experimental data on the photon-to-pion transition to
two-parameter models.
\begin{table}[t]
\caption{Estimates of the Gegenbauer coefficients $a_2$ and $a_4$
at the normalization scale $\mu^2_\text{SY}$ obtained by different
methods and for several pion DA models.
The designations correspond to those used in Fig.\
\protect\ref{F:DAmodels}.
Also included are the theoretical constraints derived from QCD SRs
with NLC and estimates derived from an analysis \cite{BMS02} of the
CLEO data \cite{CLEO98} using LCSRs.
The lattice measurements of \protect\cite{Lat07}, \protect\cite{Lat06},
and \protect\cite{DelDebbio05} are also shown.
[Note that the uncertainties of the coefficients $a_2$ and $a_4$
\textit{are correlated}.
Here, the rectangular limits of the fiducial ellipse
\protect\cite{BMS02,BMS03} are shown.]
 \label{tab:collage}}
\begin{ruledtabular}
 \begin{tabular}{|c|c|cc|}
DA methods/models
 & Symbols
    & $a_2(\mu^2_\text{SY})$
        & $a_4(\mu^2_\text{SY})$
             \\ \hline \hline
 \textbf{Lattice} &
   &
     &  \\
 UKQCD/RBC ~\cite{Lat07} %
             & \begin{minipage}[c]{60pt}
    $\includegraphics[width=20pt,height=10pt]{
    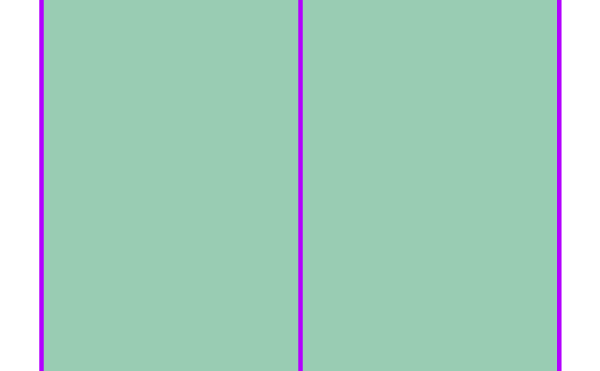}$
    \end{minipage} &$0.215\pm 0.07$
                     & --- \\
QCDSF/UKQCD ~\cite{Lat06}
 &   \begin{minipage}[c]{60pt}
    $\includegraphics[width=30pt,height=10pt]{
    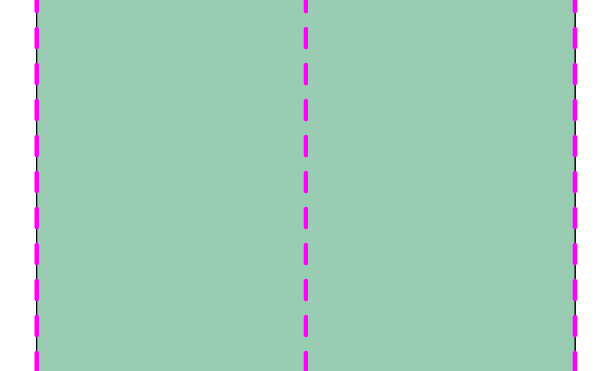}$
    \end{minipage}
    & $0.19\pm0.11$
        & ---
               \\
 ~\cite{DelDebbio05}
 &
    & $0.233\pm0.145$
        & --- \\ \hline
 \textbf{Data analysis} &
   &
     &  \\
LCSR for $F^{\gamma \pi}$, CLEO data
 &
    &
        &   \\
~SY~~ best fit~\cite{SchmYa99}
 & {\footnotesize\ding{108}}
    & $0.19\pm 0.04\pm 0.09$
        & $-0.14\pm 0.03\mp 0.09$ \\
BMS best fit~\cite{BMS02,BMS03}
 & \BluTn{\ding{58}}
    & $0.22$
        & $-0.22$
              \\
BMS~\cite{CLEO98,BMS03}($1\sigma$) &
    & [0.11, 0.328]
        & $[-0.03, -0.41]$  \\
Guo\&Liu~\cite{Guo-Liu08}($1\sigma$) &
    & [0.06, 0.14]
        & $[-0.02, -0.13]$  \\
BMS best fit~\cite{BMS05lat}&
    &
        &   \\
(twist-four via renormalons)
 & \RedTn{\ding{58}}
    & $0.31$
        & $-0.25$
              \\
Agaev  best fit~\cite{Ag05a}
 & {$\triangle$}
    & $0.23$
        & $-0.05$
               \\
LCSR for $F^{\pi}$, JLab data,
& & & \\
BK model \protect{\cite{BiKho02}}
 &
    & $0.17\pm 0.11$
        & ---
               \\ \hline
  \textbf{NLC QCD SR} &
   &
     &  \\
QCD SRs for $\langle{\xi^N}\rangle_\pi$
\protect{\cite{BMS01,BMS04kg}}
 & \begin{minipage}[c]{60pt}
    $\includegraphics[width=30pt]{
    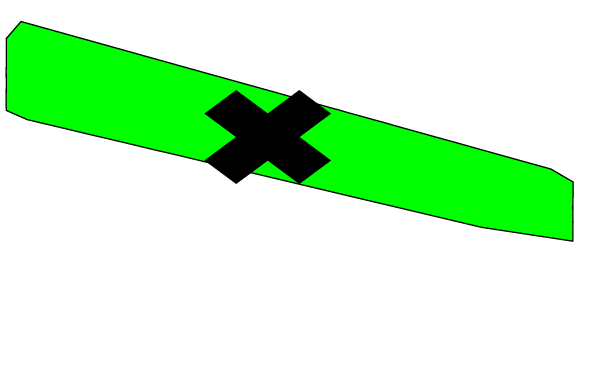}$
    \end{minipage}
     & $[0.1,~0.185]$
        & $[-0.03, -0.14]$
               \\
BMS model~\cite{BMS01,BMS04kg}
 & {\ding{54}}
    & $0.14$
        & $-0.09$
              \\ \hline
  \textbf{Instanton models} &
   &
     &  \\
ADT~\cite{ADT00}
 & ~{\footnotesize\ding{115}} 
    & $0.034$
        & $-0.027$
              \\
PPRWG~\protect{\cite{PPRWG99}}
 & {\ding{73}}  
    & $0.03$
        & $0.005$
               \\
PR~\protect{\cite{PR01}}
 & \BluTn{\ding{70}} 
    & $0.06$
        & $-0.01$
\\
NK~\protect{\cite{Kim06}}
 \footnote{\footnotesize This instanton model improves the works in \cite{PPRWG99,PR01}
 and turns out to be inside the $2\sigma$ error ellipse of the CLEO data.}
 & 
    & $0.119$
        & $0.014$
               \\ \hline
  \textbf{Models} &
   &
     &  \\
As (Asymptotic)
 & \RedTn{\ding{117}}
    & 0
        & 0
              \\
BZ~\protect{\cite{BZ05}}
 &  \RedTn{\footnotesize\ding{115}}
    & $0.08$
        & $-0.01$
              \\
CZ~\protect{\cite{CZ84}}
 & \RedTn{\footnotesize\ding{110}}
    & $0.40$
        & 0
               \\
BF~\protect{\cite{BF89}}
 & \RedTn{{\footnotesize\ding{116}}} 
    & $0.31$
        & 0.15
               \\
AdS/QCD~\protect{\cite{BT07}}
  &
    &
      & \\
$(a_2,a_4)$ projection
 & \RedTn{{\footnotesize\ding{108}}} 
    & $0.1$
        & 0.035
               \\
\end{tabular}
 \end{ruledtabular}
  \end{table}
Such a data analysis of the CLEO measurements was first performed by
Schmedding and Yakovlev (SY) \cite{SchmYa99} within the framework of LCSR.
In a subsequent series of papers
\cite{BMS02,BMS03,BMS04kg,BMS05lat,BMPS07,Ste08} (nicknamed BMS), this
type of data processing was further pursued with results confirming the
previous SY findings while also improving the theoretical accuracy.
These results are displayed in Fig.\ \ref{F:DAmodels} with details
being provided in Table \ref{tab:collage}.
In this graphics (left panel), the slanted shaded (green) rectangle
represents the BMS DA ``bunch'' from the NLC approach together with the
middle point {\footnotesize\ding{54}}, which corresponds to the BMS
model with the coefficients
$a_2^\text{BMS}(\mu^2_\text{SY})=+0.14,
a_4^\text{BMS}(\mu^2_\text{SY})=-0.09$.
The CLEO-data constraints are shown in the form of error ellipses:
$1\sigma$ (thick solid green line), $2\sigma$ (solid blue line),
and $3\sigma$ (dashed-dotted red line).
The range of values of $a_2$, determined recently on the lattice
by two independent Collaborations, is denoted by vertical
dashed lines \protect\cite{Lat06} and solid ones \cite{Lat07}.
All constraints and predictions shown have been evolved to the scale
$\mu^2_{\rm SY}$ using the NLO ERBL evolution equation.

In the right panel of Fig.\ \ref{F:DAmodels} we show the effect
of the twist-four contribution to the pion DA, taken into account via
the renormalon approach of \cite{BGG04}, in comparison with the
standard one, which is based on the asymptotic DAs of
twist four \cite{Kho99}, namely,
\ba
  \varphi^{(4)}_{\rm as}(x,\mu^2)
=
  \frac{80}{3}\delta^2(\mu^2)\, x^2 (1-x)^2\, .
\ea
Here $\delta^2(1{\rm GeV}^{2}) = (0.19\pm 0.02)~{\rm GeV}^{2}$.
This estimate has been obtained in \cite{BMS02} under the
assumption that
$
 \lambda^2_q
\equiv
 \langle
        \bar{q} \left(i g\,\sigma_{\mu \nu}G^{\mu \nu} \right) q
 \rangle
 /(2\langle \bar{q} q \rangle)
=
  0.4\pm 0.05$~GeV$^{2}$ (see Appendix A in \cite{BMS02}).
A full-fledged analysis of the renormalon-model corrections has been
given in \cite{BMS05lat}---see also \cite{Ag05b}.
The net effect is to shift the whole $1\sigma$ error ellipse (broken
contour in Fig.\ \ref{F:DAmodels}) further away from the asymptotic
pion DA and hence to higher values of $a_2$ (observe the shift of the
best-fit point to its new position denoted by \RedTn{\ding{58}}).
The original $1\sigma$ error ellipse (solid contour) is also shown
in this figure and one appreciates that the variation of the
twist-four term can have substantial influence on the transition form
factor.

Let us conclude this subsection with the following remarks:
(i) The most striking message of Fig.\ \ref{F:DAmodels} (to be read
in conjunction with Table \ref{tab:collage}) is that the CLEO $1\sigma$
error ellipse, the two recent lattice calculations, and the fiducial
region of pion DAs extracted from nonlocal QCD sum rules all have a
common region of validity.
Moreover, the BMS model DA {\ding{54}} is entirely within this area.
(ii) In contrast, the asymptotic pion DA {\RedTn{\ding{117}}} is
outside the 3$\sigma$ ellipse, while the CZ pion DA
{\RedTn{\footnotesize\ding{110}}}, as well as the BF model
{\RedTn{\footnotesize\ding{116}}, are outside the 4$\sigma$ error
ellipse of the CLEO data.
(iii) All models inside the ``rhombus'', determined in \cite{BZ05},
are also more or less outside the $1\sigma$ error ellipse.
This fact gains more weight in view of the reduced interval of $a_2$
computed on the lattice \cite{Lat07}
(in the form of $\langle \xi^2 \rangle_\pi$)
which seems to exclude all these models as well.
(iv) Analogous considerations for the moment
$\langle \xi^4 \rangle_\pi$ have been discussed in \cite{Ste08}.

\subsection{Phenomenological input of the light-cone sum rules}
\label{subsec:pheno-input}
The next task concerns the model of the resonances entering the
dispersion integral for the transition form factor.
As we mentioned before, we are going to refine the simple
$\delta$-function ansatz by taking into account a finite width in
terms of the Breit-Wigner model.
To this end, we replace $\rho^{\rm h}$ in
\begin{eqnarray}
&& \rho^{\rm phen}(s,Q^2)= \theta(s_0-s)\rho^{\rm h}(s,Q^2)+
    \theta(s-s_0) N_{\rm T}~\rho(Q^2,s) \label{eq:ro}
\end{eqnarray}
by
\begin{eqnarray}
&&     \rho^{\rm h}(s,Q^2)=\frac1{\sqrt{2}\pi}\sum_{V=\rho,\omega}
     \frac{m_V\Gamma_V }{\left( m_V^2-s\right)^2
     +m_V^2 \Gamma_V^2}\ f_V F^{\gamma^* V \pi }(Q^2) \label{eq:mes}
\end{eqnarray}
with $~f_{\omega} \simeq f_{\rho}/3, \mbox{and}
     ~~F^{\gamma^*\omega\pi}\simeq 3 F^{\gamma^* \rho\pi}$.
Then, we trade the simple $\delta$-function resonance model
for the Breit-Wigner ansatz \cite{Kho99}, given by
\be
\label{Deltaansatz}
  \delta(s-m^2_V) \to \Delta_V(s)
\equiv
  \frac1{\pi}\frac{m_V\Gamma_V }{\left( m_V^2-s\right)^2
  +m_V^2 \Gamma_V^2} \ ,
\ee
also taking into account the difference in the masses and widths of
the
$\rho$ and the $\omega$ vector mesons,
$m_{\rho} = 0.7693$~GeV, $m_{\omega} = 0.7826$~GeV, and
$\Gamma_{\rho} = 0.1502$~GeV, $\Gamma_{\omega} = 0.00844$~GeV,
respectively.
To continue, in order to obtain $ F^{\gamma^* \rho\pi}$, we appeal to
the duality between the hadronic part of the spectral density and its
perturbative counterpart that we express via
\begin{equation}
\label{eq:LCSR}
  \int\limits_{4m_{\pi}^2}^{s_0}\!\!\frac{\rho^{\rm h}(s,Q^2)}{s+q^2}ds
=
  N_{\rm T} \int\limits_{0}^{s_0}\!\!\frac{\rho(Q^2,s)}{s+q^2}\, ds \ .
\end{equation}
The main effect of using Eq.\ (\ref{Deltaansatz}) on the LHS of
Eq.\ (\ref{eq:LCSR}) is a slight suppression of the integral relative
to the outcome of the simple $\delta$--function ansatz.

After the Borel transformation  of Eq.\ (\ref{eq:LCSR}), one arrives
at an expression for $F^{\gamma^* \rho\pi}$ in terms of $V$
[cf.\ (\ref{eq:Fgammarhopi})], notably,
\begin{subequations}
\label{eq:GRP-BW}
\ba \label{eq:Bw-ffBorel}
  F^{\gamma^* \rho\pi}(Q^2)
&=&
  k^{-1}\frac{f_\pi}{3 f_{\rho}} V(Q^2, M^2) \ , \\
  k
&=&
  \int\limits_{4m_{\pi}^2}^{s_0}\!
  \frac{\Delta_\rho(s)+\Delta_\omega(s)}{2}
  \exp\left(\frac{m_\rho^2}{M^2}-\frac{s}{M^2}\right) ds \ ,
\ea
\end{subequations}
where $k\approx 0.932$ weakly depends on the Borel parameter $M^2$
in a region containing the standard scale
$M^2 \simeq 0.7$~GeV$^2$.
Therefore, the Breit-Wigner ansatz for the resonances
(abbreviated in what follows by BW) supplies a factor
$k^{-1}> 1$, entailing an increase of the value of the form factor
as compared to the simple $\delta$-function ansatz.
A similar enhancement effect in analogy to Eq.\ (\ref{eq:srggpi2})
applies to the total form factor
$F_\text{LCSR}^{\gamma^*\gamma\pi}$,
leading to
\begin{subequations}
\ba
  F_\text{LCSR}^{\gamma^*\gamma\pi}(Q^2)
&=&  N_T \left\{\frac{k_1}{k}\frac{1}{m^2_\rho} V(Q^2, M^2)
  + \frac{1}{Q^2} H(Q^2) \right\}\, ,
\label{eq:srggpi2BW} \\
  k_1&=& \int\limits_{4m_{\pi}^2}^{s_0}\! \frac{\Delta_\rho(s)
  +\Delta_\omega(s)}{2}
  m_{\rho}^2 \frac{ds}{s}\, ,
\ea \label{eq:k_1}
\end{subequations}
$\!\!\!\!$ where $k_1 \approx 0.984$.
Crudely speaking, the net enhancement amounts to 2-4\% in that
momentum region, where the resonance part prevails over the
pointlike one.

\subsection{NNLO effects on
            $\mathbf{Q^2 F^{\gamma^{*}\gamma\pi}}$---integral
            characteristics}
\label{subsec:gamma-gamma-pi-data-comp}
Having set up the framework for calculating the pion-photon transition
form factor, including also its nonperturbative ingredients, we now
turn our attention to the specific effects due to the NNLO$_\beta$
radiative corrections according to Eq.\ (\ref{eq:generNNLO}).

Being interested mainly in the magnitude of the form factor, it is
actually sufficient to include only the $\psi_0$ term.
An analysis including higher harmonics will be presented in a future
publication.
\begin{figure}[ht]
 \centerline{\includegraphics[width=0.65\textwidth]{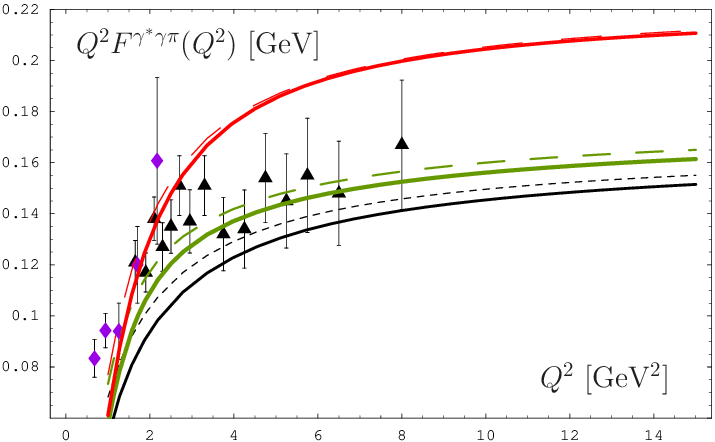}~}
 \vspace*{-1mm}
 \caption{\footnotesize  Theoretical predictions from LCSR for
   the form factor $Q^2 F^{\gamma^{*}\gamma\pi}$ with the
   NNLO$_\beta$ corrections (together with the BW resonance
   model) included (solid lines) and without them (dashed lines).
   All predictions shown are evaluated with the twist-four parameter
   value
   $\delta_{\rm Tw-4}^2=0.19$~GeV$^2$~\protect\cite{BMS02,BMS03}.
   They correspond to selected pion DAs. These are:
   CZ model---upper (red) line \protect\cite{CZ84},
   BMS-model---middle (green) line \protect\cite{BMS01},
   and As DA---lower (black) line.
   For comparison, the corresponding predictions for each pion DA model
   without these corrections are displayed as dashed lines.
   The experimental data shown are from the CELLO (diamonds,
   \protect\cite{CELLO91}) and the CLEO (triangles,
   \protect\cite{CLEO98}) Collaborations.
   }
 \label{fig:Fig6}
\end{figure}

The final results for the photon-to-pion transition form factor,
including the NNLO$_\beta$ corrections, within the improved LCSR
approach are displayed in Figs.\ \ref{fig:Fig6} and \ref{fig:Fig7}.
The presented results have been obtained along the lines followed in
our previous dedicated and detailed works in
\cite{BMS02,BMS03,BMS05lat} and will not be repeated here.
Recalling the features of the pion DAs discussed in Subsec.\
\ref{subsec:nonperturbative}, we reduce our discussion to only three
models: As (lower black line), CZ (upper red line),
and BMS (green line in the middle).
The solid lines represent the scaled form factor with the NNLO$_\beta$
and the BW--ansatz corrections included, whereas the dashed lines (with
the corresponding color for each model) show the result without these
corrections.
These theoretical predictions are displayed in the background of the
experimental data from the CLEO \cite{CLEO98} (triangles) and the CELLO
(diamonds) \cite{CELLO91} Collaborations.

The main lesson from these figures is that the effect on
$Q^2 F^{\gamma^{*}\gamma\pi}$, induced by the radiative corrections,
amounts to about ($-10 \div -5$)\% (see Sec.\ \ref{sec:FF-results}),
whereas that caused by the use of the more realistic BW-resonance
ansatz provides a small growth of approximately ($+4 \div +2$)\%.
Combining both effects, results into a net reduction of the magnitude
to within 7\% at small $Q^2 \simeq 2$~GeV$^2$.
In the case of the As DA, the size of this reduction rapidly drops to
2.5\% for $Q^2\geq \mu_{\rm SY}^2$ , as one may appreciate by
comparing the solid line (NNLO$_\beta$) with the dashed line (NLO) at
the bottom of Fig. \ref{fig:Fig6}.
These results have been obtained with the spectral density
$\rho^{(2 \beta)}_{0}(x;s,Q^2,Q^2)
=
  C_{\rm F} \bar{R}^{(2)}_0(x,\bar{x}/x)
$,
where the further evaluation goes along Eqs.\
(\ref{eq:C1})--(\ref{eq:Tbeta}) for the $\psi_0$--harmonic.
On the other hand, the transition form factor is mainly determined
via the inverse moment \cite{BMS04kg}, which belongs to the
\emph{integral characteristics} of the pion DA, i.e.,
\begin{equation}
  \langle x^{-1}\rangle_{\pi}
\equiv
  \int_{0}^{1}dx \frac{\varphi_\pi(x)}{x}
=
 3(1+a_2+a_4+\ldots) \ ,
\label{eq:inv-mom}
\end{equation}
that can be easily understood from the expression for the $H$
part in Eq.\ (\ref{eq:Hrho-x}) setting $Q^2 \gg s_0,~x_0 \to 1$.
This means that for pion DAs, like the BMS one, which have the
particular property $a_2 \sim - a_4$ (see the anti-diagonal
in Fig.\ \ref{F:DAmodels}), the corrections associated with the higher
harmonics $\psi_2$ and $\psi_4$ mutually cancel, leaving only the
correction due to $\psi_0$.
The same conclusion can be drawn for all pion DAs belonging to the
BMS ``bunch''---shown as a green strip in Fig.\
\ref{fig:Fig7}---(consult Sec.\ \ref{subsec:nonperturbative} for
further explanations).\footnote{Note that the width of the BMS strip is
somewhat narrower compared to our previous results in \cite{BMS04kg}
because here we use a smaller range for the $\delta^2$ uncertainties
in the twist-four contribution.}
This cancelation effect is absent for the CZ DA, as one observes by
comparing the solid (red) line (NNLO$_\beta$) with the dashed (red)
line (NLO) on the top of both these figures.
The reason is that the CZ DA has only a large $a_2$ coefficient,
while all other coefficients are zero.
Hence, we cannot estimate the size of the NNLO correction using only
the $\psi_0$ part of the NNLO contribution, but we have to include the
corrections for higher harmonics as well.

\begin{figure}[h]
 \centerline{
             \includegraphics[width=0.66\textwidth]{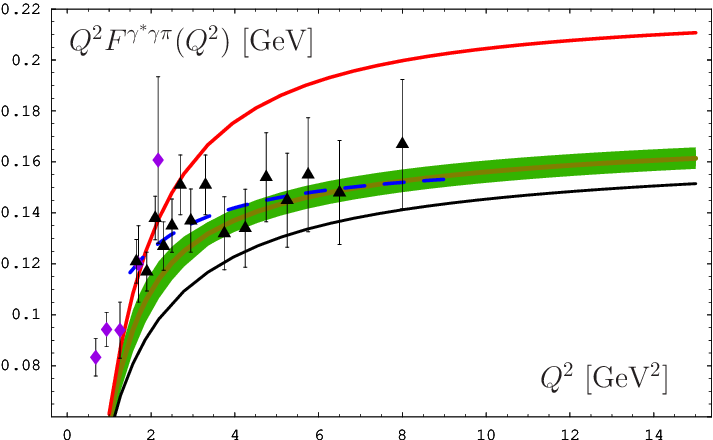}
 }
\vspace*{-1mm}
 \caption{\footnotesize
 The same designations as in the previous figure apply.
 The shaded (green) strip denotes the form-factor predictions derived
 with the help of the BMS ``bunch'' \protect\cite{BMS01}.
 The dashed blue line represents a dipole-form interpolation
 formula \protect\cite{CLEO98} of the CLEO data.
 }
 \label{fig:Fig7}
\end{figure}

We observe from Figs.\ \ref{fig:Fig6} and \ref{fig:Fig7} that
the improved theoretical predictions which contain the discussed
effects fail to describe both sets of the experimental data at low
$Q^2$, say, below 4~GeV$^2$.
Actually, the data can be grouped into three regions:\\
(i) The low $Q^2$ region is mainly covered by the CELLO data and
extends to momenta up to approximately 1~GeV$^2$.
This region, where hadronization is immanent, is virtually inaccessible
to the methods used in our analysis---let us call it therefore the
\emph{unknowable} regime.\\
(ii) There is some intermediate momentum region between 1 and
3~GeV$^2$, where the existing data are underestimated by our
theoretical predictions.
In contrast to the previous case, here we can figure out what the
origin of this drawback may be.
Hence, let us call this intermediate domain the \emph{unknown} regime,
because the missing contributions can, in principle, be estimated. \\
(iii) Finally, above about 3~GeV$^2$, the agreement between the CLEO
data and the shaded BMS strip is fairly good, as one also realizes by
comparing these data points with the dipole fit (dashed blue line)
used by the CLEO Collaboration.
In fact, the dipole curve and the prediction due to the BMS model
(solid curve within the BMS strip) almost coincide in this $Q^2$
region.
Therefore, it is safe to claim that this high-momentum domain is
well-reproduced by our techniques---hence the term \emph{known}
regime.\footnote{However, one may face the challenge posed by the
recently released BaBar data \cite{BaBar09} exactly in this
momentum-transfer region---see Note Added below.}

Let us now discuss the reasons for the discrepancy in the
\emph{unknown} regime.
\begin{itemize}
\item One reason for the observed discrepancy may be traced to
the fact that we used for the evolution of the twist-four
contribution only the one-loop anomalous dimension.
Considering the evolution effect at the two-loop level, could
potentially reduce its size, thus rendering the reduction of the
transition form factor less severe.
\item The uncalculated remnant of the NNLO contribution could
eventually enhance its total magnitude---should this part appear with
the opposite sign with respect to the calculated
$\beta$-function part.
\item Still another source of uncertainty lies with the value
of $\delta^2$ which controls the size of the twist-four contribution.
A smaller value of this parameter would entail less suppression of the
form factor, given that this nonperturbative contribution has a
negative sign, just like the calculated NNLO$_\beta$ contribution.
Here there is a subtlety.
If one assumes a smaller value of $\delta^2$, then this would
automatically mean that also $\lambda_q^2$ should assume a smaller
value because $\delta^2\approx \lambda_q^2/2$, i.e., these two
nonperturbative parameters work synergistically.
But a decrease of the latter parameter would yield to an enhancement
of the leading-twist contribution, meaning that the whole BMS strip
(made up on the basis of the BMS ``bunch'' \cite{BMS01,BMS04kg})
would move somewhat upward as a whole.
As a result, one would obtain an increase of the transition form factor
exactly in that regime between 2 and 4~GeV$^2$, providing a better
agreement with the CLEO and CELLO data.
\end{itemize}

\subsection{NNLO effects on
            $\mathbf{Q^{4}F^{\gamma^{*}\rho\pi}(Q^2)}$---differential
            characteristics}
\label{subsec:gamma-rho-pi}
The scaled form factor
$Q^{4}F^{\gamma^{*}\rho\pi}(Q^2)$, obtained in the framework of LCSRs,
see Eq.\ (\ref{eq:GRP-BW}), is determined by the hadronic part $V$,
defined in Eq.\ (\ref{eq:Vrho-x}).
At moderate $Q^2$ around the scale $\mu^2_{\text{SY}}$, $V$ is mainly
formed by the $\varphi_\pi$--dependent leading-twist contribution and
to a lesser extent by the twist-four one.
The combined effect of the BW-ansatz and the NNLO$_\beta$ corrections
is rather important and amounts to approx.\ $-10\%$ in the region
limited from above by the scale $\mu^2_{\text{SY}}$.
For still higher momentum values, and up to
$Q^2 \approx 15$~GeV$^2$, it reaches the level of $+9\%$---see left
panel of Fig.\ 8.
More specifically, the NNLO$_\beta$ corrections reduce the value of
the form factor by an amount of approx. 20\%, starting at
$Q^2=2$~GeV$^2$,
and become significantly smaller at the end of this region
(see dashed line in Fig.\ 8, left panel), whereas the use of the
BW-ansatz leads to an overall increase of 8\%.
The ratio of the combined effect of the NNLO$_\beta$ and the
BW-ansatz contributions, relative to the total transition form factor,
is illustrated in Fig.\ \ref{F:Fig8} in terms of a solid red line,
whereas the normalized contribution of the NNLO$_\beta$ term alone is
depicted in the same figure by the dashed blue line.

The form factor $Q^{4}F^{\gamma^{*}\rho\pi}(Q^2)$ is presented in
Fig.\ 8 (right panel) using different pion DAs.
We observe from this figure that this process can actually be
used to discriminate among different pion DAs of twist two, as first
pointed out by Khodjamirian in \cite{Kho99} because it is
sensitive to the particular shape of the pion DA.
Especially the momentum region between 2~GeV$^2$ and 8~GeV$^2$ seems
to be particularly convenient for this task because (i) the
obtained predictions are clearly distinguishable and (ii) because it
corresponds to the range probed already by the CLEO experiment
for the pion-photon transition \cite{CLEO98}.
The measurement of \emph{both} transition form factors in this
momentum region would provide definitive clues for the underlying pion
DA.
\begin{figure}[ht]
 \centerline{\includegraphics[width=0.48\textwidth]{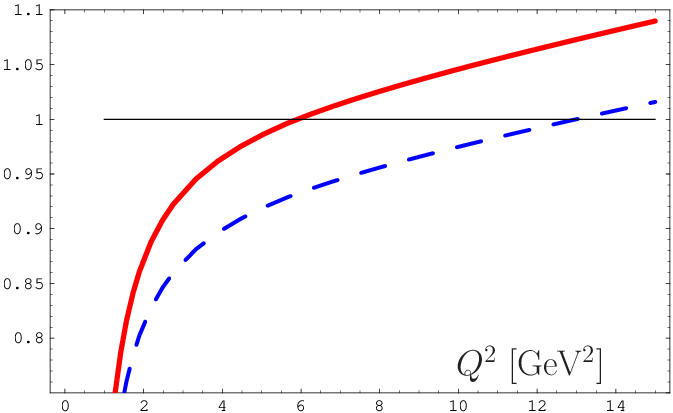}~~~~~~
             \includegraphics[width=0.48\textwidth]{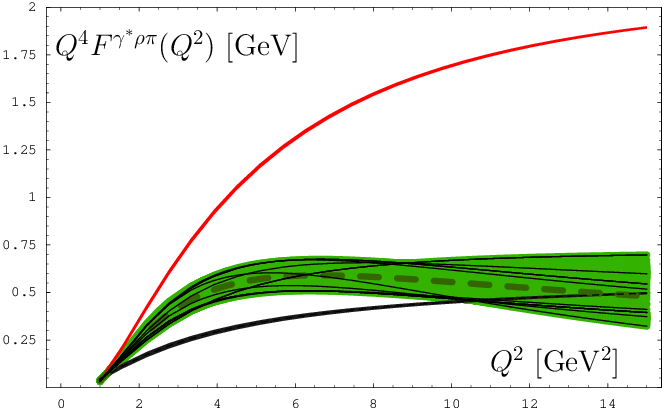}
 }
 \vspace*{-0mm}
 \caption{\footnotesize Left: The upper solid curve shows the combined
 contribution of the NNLO$_\beta$ and the BW-ansatz.
 The dashed line below it represents the normalized contribution of the
 NNLO$_\beta$ only.
 Right: Predictions for $Q^4 F^{\gamma\rho \pi}(Q^2)$ for different
 pion DAs:
 the top solid (red) line denotes the CZ DA, the middle (green)
 strip represents the BMS ``bunch''---including the BMS model DA as a
 dashed line---, and the lower solid (black) line gives the
 result for the As DA up to 15~GeV$^2$.
 The other lines shown inside the BMS strip are explained in the
 text.}
 \label{F:Fig8}
\end{figure}

Let us now make some detailed remarks on the structure of the
BMS strip, obtained from the ``bunch'' of pion DAs following
from  NLC QCD sum rules.
The first observation from Fig.\ 8 (right panel) is that the BMS strip
(which includes the BMS model: long-dashed line)
\cite{BMS01} crosses around 11~GeV$^2$ the prediction obtained with the
As DA.
We argue that this effect can be traced back to the
\emph{differential} pion characteristics of the pion DA, viz.,
$\displaystyle
\varphi_{\pi}'(0)\equiv \frac{d}{dx} \varphi_{\pi}(x)|_{x=0}
$.
The usefulness of the slope of the pion DA was first discussed
in \cite{BH94} in connection with the pion's electromagnetic form
factor.
In our case, it is instructive to recall the
definition of $V$ in Eq.\ (\ref{eq:Vrho-x}), from which one appreciates
that the lower limit of the dispersion integral $x_0$ tends to the
upper one, $1$, when $Q^2 \gg s_0$.
Therefore, near the upper integration limit, this integral is
determined by the endpoint behavior of $\bar{\rho}(x)|_{x\sim 1}$,
which in LO is proportional to
$\varphi_{\pi}'(0)$ [cf.\ Eq.\ (\ref{eq:rho-0})], entailing
for the form factor a behavior like
$\Ds \sim \varphi_{\pi}'(0)\frac{s_0^2}{\left(Q^2+s_0\right)^2}$.
This becomes visible in the vicinity of the upper
limit 1, say, for a value $ 0.1= s_0/(s_0+Q_*^2)$,
which for the BMS model corresponds to $Q_{*}^2\approx 14$~GeV$^2$,
while for the upper part of the strip the value of $Q_*^2$ is much
larger.
Thus, it becomes possible to investigate the endpoint behavior of
different pion DAs---mentioned also in \cite{BF89,Kho99}---in terms
of this transition form factor by probing the slope of the pion DA
at the origin.}

To make these statements more quantitative, let us express
the slope of the pion DA in terms of the Gegenbauer coefficients to
obtain
\begin{eqnarray}
\label{eq:deriv}
  \varphi_{\pi}'(0;\mu^2)
=
  6\left(1+\sum_{m=2,4,\ldots}~(m+1)(m+2)/2 \ a_{m}(\mu^2) \right) \ .
\end{eqnarray}
Then, we get the following results:
\begin{eqnarray}
 \label{eq:endpointmod}
\varphi_{\pi}'(0;Q_{*}^2) =  \left\{
\begin{array}{l}
   17.4~~\text{[CZ]}\\
    ~3.7~~\text{[BMS]}\\
    ~5.8~~\text{[As]} \ .
 \end{array}
   \right.
 \end{eqnarray}
These simple estimates are qualitatively responsible for the
associated form-factor values at $Q_{*}^2$ (slopes and altitudes) in
Fig.\ \ref{F:Fig8} (right panel).
It turns out that for the CZ DA the value of the form factor
is approximately 3 to 4 times larger than the one for the As DA,
while for the BMS DA it is smaller or quite close to it
[cf.\ Eq.\ (\ref{eq:endpointmod})].
Note that this feature appears to be opposite to the
$Q^{2}F^{\gamma^{*}\gamma\pi}(Q^2)$ form factor
that mainly depends on the inverse moment \cite{BMS05lat}
$
 \langle x^{-1}\rangle_{\pi}
$,
the latter being an \emph{integral} pion characteristic
\cite{BMS01, BMS03}.
Indeed, one sees from Fig.\ \ref{fig:Fig7} that the results
for $Q^{2}F^{\gamma^{*}\gamma\pi}(Q^2)$ (green-shaded strip) are
always larger than the predictions obtained from the asymptotic DA.
Hence, $Q^{4}F^{\gamma^{*}\rho\pi}(Q^2)$ can provide complementary
information about the pion DA and help discriminate among various
proposed pion DA models.

Going beyond the leading order of $\bar{\rho}$, a new behavior near
the endpoints emerges due to the hard-gluon exchange leading to
a $1/Q^2$--behavior of the form factor in the far asymptotic
domain.
This is, because in NLO, the expression $\bar{\rho}^{(1)}$ in
Eqs.\ (\ref{eq:bar-rho}) and (\ref{eq:rho1-n})
contains a term proportional to $\Ds -\ln^2(\bar{x}/x)$ that
accumulates this effect.

A second issue related to the BMS strip in the right panel of
Fig.\ \ref{F:Fig8}, which deserves to be considered as well, concerns
the fine structure of its envelopes.
Indeed, close inspection reveals that the upper envelope has a dip at
$Q^2\approx 9$~GeV$^2$, whereas the analogous irregularity of the
lower envelope is much less pronounced and appears at a scale close
to 10.5~GeV$^2$.
The origin of these irregularities might be related to the crossing of
the predictions obtained from different DAs inside the BMS ``bunch''
that have different initial values of the Gegenbauer coefficients $a_2$
and $a_4$ (exemplified by the intersecting lines inside the BMS strip).
When it happens that such a crossing point lies just near the upper or
lower boundary of the BMS strip, the corresponding envelope is ``bent''
inwards.
Physically, this means that the range of the calculated uncertainties
is somewhat smaller there relative to the regions where no crossing
points appear close to the boundaries.
On the other hand, we cannot exclude that these irregularities are
a spurious effect created by the algorithm we used to obtain our
predictions.
In any case, they have no influence on our analysis.

\section{Conclusions}
\label{sec:concl}
Results have been presented for the photon-to-pion and the
$\rho$-meson-to-pion transition form factors using light-cone sum
rules and including those NNLO contributions which are proportional
to the $\beta$ function.
Let us summarize our main findings.

\begin{enumerate}
\item
The spectral density of the LCSR at the NLO level was systematically
constructed for any Gegenbauer harmonic of order $n$ and was presented
in compact form in Eqs.\ (\ref{eq:rho1}) and (\ref{eq:rho1-n}).
\item
Using the NNLO hard-scattering amplitude $T_{\beta}^{(2)}$,
proportional to the $\beta$ function and calculated before in
\cite{MMP02}, we derived all necessary ingredients of the NNLO spectral
density $\rho^{(2)}_{n}$ and analyzed $\rho_{0}^{(2)}$ for the $\psi_0$
harmonic via Eqs.\ (\ref{eq:C1})--(\ref{eq:Tbeta}).
These quantities enter the dispersion integral which determines the
structure of the form factors
$F^{\gamma^{*}\rho\pi}$ and $F^{\gamma^*\gamma\pi}$
within the light-cone sum-rule approach.
\item
Predictions for the form factor $F^{\gamma^*\gamma\pi}$ were
presented which include the NLO and the $\beta$-part of the NNLO
radiative correction, an improved resonance contribution, and also
estimates of the uncertainties stemming from the twist-four
contributions.
The first contribution has a negative sign and amounts to a
reduction of up to 10\%, whereas the Breit-Wigner ansatz provides a
small enhancement below 5\%,
and the twist-four contributions can yield
to a small enhancement by adopting a smaller value of the parameter
$\delta^2$ which controls their size.
This, however, would entail a parallel enhancement of the leading-order
contribution to $F^{\gamma^*\gamma\pi}$ via the parameter
$\lambda_q^2$, because these two parameters are correlated
($\delta^2\approx \lambda_q^2/2$).
Comparison with the CLEO and the CELLO data shows that at 2 GeV$^2$,
and below, the calculated form factor falls slightly short compared
to the data, leaving room for additional soft and higher perturbative
contributions that are unknown at present.
\item
We have also given predictions for the $F^{\gamma^{*}\rho\pi}$
form factor using different pion distribution amplitudes including the
asymptotic one, the whole set of pion DAs derived from nonlocal
QCD sum rules, and the CZ model.
In this case, the combined effect of the negative NNLO$_\beta$
radiative corrections and the use of the Breit-Wigner ansatz lead to
a net reduction of the form-factor magnitude varying between $-10\%$,
below the scale $\mu^2_{\text{SY}}$, and growing to the level of
$+9\%$ at $Q^2 \approx 15$~GeV$^2$, whereas the twist-four
contributions amount to $-11\%$ even up to the scale $15$~GeV$^2$.
Once there will be experimental data for this reaction, it will
provide an additional and useful tool to discriminate among various
pion DAs---especially between those of the CZ type that receive strong
endpoint enhancement on the one hand and such with their
endpoints being suppressed---like the BMS one---on the other.
Moreover, because this form factor is sensitive to the
\emph{differential characteristics} of the pion DA---expressed via
$\varphi_{\pi}'(0)$---one can use the endpoint behavior of the
underlying pion DAs as an additional adjudicator in selecting the
optimal pion DA.
\end{enumerate}

To conclude, we have extended the analysis of the pion-to-photon
transition process to the NNLO level in a systematic though
partial way.
Nevertheless, our analysis provides the possibility to estimate
the size of the associated form factors in that region of momenta which
is accessible to present measurements.
It will be interesting to pursue and complete this sort of
calculation in the future by including the whole NNLO contribution.

\begin{acknowledgments}
We would like to thank Alexander Bakulev for collaboration, Dieter
M\"uller for useful discussions, and Micha{\l} Prasza\l{}owicz for
communications.
One of us (M.V.S.) is indebted to Prof.\ Klaus Goeke
for the warm hospitality at Bochum University, where part of this
investigation was carried out.
This work was partially supported by the Heisenberg--Landau
Program (grant 2009), the Deutsche Forschungsgemeinschaft (DFG)
under grant 436 RUS 113/881/0, and the RFBR (grant 09-02-01149).

We are grateful to Alexander Pimikov and Alexander Bakulev for their
valuable help in verifying the Mathematica codes and some results 
of our analysis, an action triggered by the error in Eq. (3.12) 
pointed out by Agaev \textit{et al.}, in Phys. Rev. D 83 (2011) 054020.
The errors found have been corrected in this version and are boldfaced for
an easier identification.
\end{acknowledgments}

\section*{Note Added}
\label{sec:note}
After completion of this work we became aware
(thanks to Dieter M\"uller and Maxim Polyakov)
of new preliminary data of the BaBar
Collaboration\footnote{Talk presented by Selina Li on behalf of the
BaBar Collaboration at Photon 2009, Hamburg, Germany, 11-14th May
2009.}
on two-photon-induced processes in the momentum range
$4<Q^2<40$~GeV$^2$ which show above 10~GeV$^2$ ``a power-law growth
behavior that contradicts most models for the pion DA''.
Meanwhile, these data have been officially released \cite{BaBar09}.

Taking the high-$Q^2$ data points of BaBar at face value, one might
even come to the conclusion that they are incompatible with QCD per se
because the indicated power-law enhancement at large momentum values
cannot be explained by any known QCD effect, like higher-order
radiative corrections or contributions due to higher-twist effects.
Our presented analysis serves to prove that the inclusion of the main
part of the NNLO radiative corrections provides suppression---not
enhancement---at the 10\% level, while the twist-four contribution
gives a small enhancement of a few percent in the CLEO region.
The expected size of the uncalculated NNLO remainder---even if it
should have a positive sign---is not expected to exceed the few-percent
level ($<10\%$).
An enhancement of the twist-four contribution due to two-loop evolution
is possible but it is expected to be of the order of a few percent as
well.
The size of all these QCD corrections is far less than indicated by the
high-$Q^2$ BaBar data, so that the observed enhancement at high $Q^2$
cannot be explained by higher-order perturbative QCD and power
corrections.

\begin{figure}[h]
 \centerline{
             \includegraphics[width=0.7\textwidth]{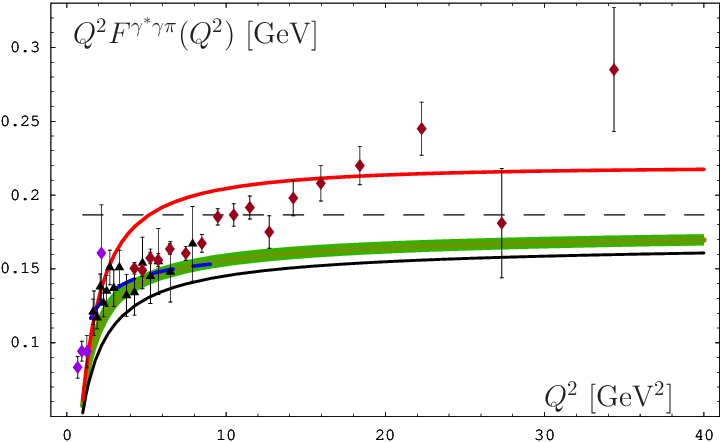}
}
\vspace*{-1mm}
 \caption{\footnotesize
 Predictions for $Q^2F^{\gamma^{*}\gamma\pi}(Q^2)$ derived from
 three different pion DA models: Asymptotic, BMS, and CZ.
 The same designations as in Fig.\ \ref{fig:Fig7} are used.
 The momentum-transfer range is extended to 40~GeV$^2$ in order to
 include the new BaBar data \protect\cite{BaBar09} (shown as thick
 bullets with error bars).
 The displayed theoretical results include the NNLO$_{\beta}$
 radiative corrections and the twist-four contributions.
 The horizontal dashed line represents the asymptotic QCD
 prediction $\sqrt{2}f_\pi$.
 }
 \label{fig:Fig9}
\end{figure}

Note that endpoint-enhanced pion DAs, like the CZ model, are also
in conflict with these data, despite opposite claims in \cite{BaBar09},
because the corresponding $Q^2F^{\gamma^{*}\gamma\pi}(Q^2)$ prediction
(see Fig.\ \ref{fig:Fig9}) scales with $Q^2$ (analogously to all other
pion DA models in the convolution scheme) above approximately
15~GeV$^2$, in sharp contrast with the significant growth of the BaBar
data in this momentum range.
In this context, we find it remarkable that the BaBar data points in
the momentum range already probed by the CLEO Collaboration are
compatible with the CLEO dipole fit and the asymptotic QCD prediction
$\sqrt{2}f_\pi$, being also within the BMS strip.
Even more significantly, two more data points---outliers---at about
14~GeV$^2$ and 27~GeV$^2$ turn out to be just on the upper boundary of
the BMS strip and, hence, in compliance with the QCD expectations.
(see Fig.\ \ref{fig:Fig9}).
Hence, one may divide the BaBar data into two branches: one containing
the data points in the CLEO region plus the two outliers, the other
consisting of the remaining 10 high-$Q^2$ data points.
The first branch supports the QCD predictions with NNLO radiative
corrections and twist-four contributions.
The other branch is in clear conflict with the convolution scheme of
QCD.
In view of this data structure, odds are that the BaBar data may
bear some intrinsic inconsistency.

\begin{table}[h]
\caption{Deviation in terms of $\bar{\chi}^2\equiv\chi^2/{\rm ndf}$
(ndf~$=$~number of degrees of freedom) of
$Q^2F^{\gamma^{*}\gamma\pi}(Q^2)$ predictions for the asymptotic
(Asy), the BMS DA, and the CZ one.
For a direct comparison with the BaBar analysis, we employ the NLO
approximation of perturbative QCD, including also twist-four
contributions.
The first column shows the results for the combined sets of the CLEO
\cite{CLEO98} and the BaBar \cite{BaBar09} data.
The second column refers only to the BaBar data, while the third column
takes into account only the last 10 high-$Q^2$ BaBar data, starting
with the data point at 10.48~GeV$^2$.
\label{tab:Babar}}
\begin{ruledtabular}
\begin{tabular}{c|c|c|c}
Pion DA models      & $\bar{\chi}^2$ CLEO and BaBar~~  & $\bar{\chi}^2$ BaBar (all data)~  & $\bar{\chi}^2$ BaBar (10 data $>10$~GeV$^2$)   \\ \hline
Asy \protect\cite{LB80,ER80} %
                    & 11.5   ~~~           & 19.2     ~~             & 19.8 ~~ \\
BMS \protect\cite{BMS01} %
                    & 4.4    ~~~           & 7.8      ~~             & 11.9 ~~ \\
CZ \protect\cite{CZ84} %
                    & 20.9   ~~~           & 36.0     ~~             & 6.0  ~~ \\
\end{tabular}
\end{ruledtabular}
\end{table}

We end this discussion with the following key observations: \\
(i) The main NNLO radiative corrections and the twist-four contributions
do not provide enhancement to $Q^2F^{\gamma^{*}\gamma\pi}(Q^2)$ in the
range of momentum transfer 10-40~GeV$^2$, exclusively covered  at present
by the BaBar experiment.
Hence, the observed behavior of $Q^2F^{\gamma^{*}\gamma\pi}(Q^2)$ growing
with $Q^2$ in this momentum region with an almost constant slope cannot be
explained within the convolution scheme of QCD.\\
(ii) Within the QCD convolution scheme, all pion DA models, which have a
convergent projection onto the eigenfunctions (Gegenbauer polynomials) of
the meson evolution equation and hence vanish at the endpoints 0 and 1, are
conflicting with the BaBar data for $Q^2F^{\gamma^{*}\gamma\pi}(Q^2)$
between 10 and 40~GeV$^2$. \\
(iii) Staying within this approach, the best agreement to the combined sets of
the CLEO and the BaBar data is still provided by the BMS-type pion DAs,
as one sees from Table \ref{tab:Babar}, first column.
Considering only the BaBar data, the deviation of the CZ DA becomes
even larger---second column in the same table.
The CZ DA is favored only when one includes in the fit the last 10
high-$Q^2$ data points, starting at 10.48~GeV$^2$ (third column in Table
\ref{tab:Babar}.)
But such a treatment of the existing data (CLEO and the 7 lower-$Q^2$ BaBar data points)
looks biased or at least unjustified.
From this discussion it becomes clear that the conclusion drawn by the BaBar
Collaboration in \cite{BaBar09} that the CZ DA is in agreement with their
data (within the convolution scheme) is unfounded.

The bottom line: The anomalous high-$Q^2$-behavior of the pion-photon
transition form factor, found by BaBar, could potentially be of great
importance, if confirmed by independent measurements, e.g., by the BELLE
Collaboration, and would demand a new framework of analysis within QCD.
Some scenarios to explain the BaBar effect have already been proposed
\cite{Dor09,Rad09,Pol09}.

\begin{appendix}
\appendix
\section{Determining the spectral density from the discontinuity of
         the hard-scattering amplitude}
\renewcommand{\theequation}{\thesection.\arabic{equation}}
\label{app:Tstructure}\setcounter{equation}{0}
Let us first define the task ahead in a bottom-to-top approach,
starting with the LO case and finishing with the NNLO one.
First, we analyze the structure of the amplitudes themselves, then,
we construct the discontinuity for their elements in order to pair each
$T_{k}$ with the spectral density, $\rho^{(k)}$, associated with this
order of the perturbative expansion.
At the LO level we encounter just one single contribution to the
discontinuity, i.e., that one arising from $T_0\otimes f$
(where $f(x)$ is some appropriate case-dependent test function).
In NLO, $T_1\otimes f$, a new element appears:
$T_0\otimes {\rm Ln}(x)\otimes f(x)$.
Finally, at the NNLO level of the expansion, $T_{2}$, we are faced
with the discontinuity of still one more quantity, notably,
$T_0\otimes {\rm Ln^2}(x)\otimes f(x)$.
[Recall that $\rm Ln$(x) denotes the logarithm of the photon momenta
over the factorization scale---cf.\ (\ref{eq:L-n}).]

Before we continue, we mention parenthetically in this context that
the contributions to $\rho^{(k)}$, just mentioned, can be also
obtained from a generating function $R(s;\varepsilon)$, which we
display below:
\begin{eqnarray}
  \Ds R(\varepsilon,x;s,Q^2)
\equiv
  \frac{\bar{R}(\varepsilon,x;s,Q^2)}{s+Q^2}
&=&
  \frac{\theta(x>x_0)}{2(s+Q^2)}
  \left(\frac{s+Q^2}{\mu_{\rm F}}\right)^{\varepsilon}
  x^{-\varepsilon} \left(x-x_0\right)^{(\varepsilon -1)}
  \frac{\sin(\pi \varepsilon)}{\pi}
\nonumber \\
&&  + (x \to \bar{x})
\label{eq:appRtot}\\
  \rho^{(k)}(Q^2,x_0)
&=&
  \frac{1}{k~!}\left(\frac{d}{d \varepsilon}\right)^{k}
  R(\varepsilon,x;s,Q^2)\Big|_{\varepsilon=0} \, .
\label{eq:appRexp}
\end{eqnarray}
The total sum of these contributions, $\bar{R}(a_s,x;s,Q^2)$,
(following from Eq.\ (\ref{eq:appRtot})) enters the spectral density
in the form of a convolution with $V^{(0)}_+$, $R\otimes V^{(0)}_+$.
Therefore, the resummed expression does not contribute to the $\psi_0$
part by virtue of the current conservation $v(0)=0$.

\textbf{1}. From the discontinuity of $T_0$ in LO
$$
  T_0(Q^2,-s;x)
=
  \frac{1}{2}\left(\frac{1}{Q^2x -s\bar{x}} + (x \to \bar{x})\right),
$$
we have \cite{Kho99,SchmYa99}
\begin{eqnarray}
  \rho^{(0)}(Q^2,s)
& = &
  \frac{1}{\pi}\mathbf{Im}
  \Bigg[
       T_0(Q^2,-s) \otimes \varphi_{\pi} =
       \frac{1}{2(Q^2+s)}
       \int_0^1\left(\frac{1}{x-x_0}+ (x \to \bar{x})\right)
     \varphi_{\pi}(x) dx
  \Bigg]
\nonumber \\
&=&
     \frac1{2(Q^2+s)} \int_0^1\Big(\delta(x-x_0)+ (x \to \bar{x})\Big)
     \varphi_{\pi}(x) dx
\nonumber \\
& = &
  \frac{\varphi_{\pi}(x_0)}{Q^2+s}\Big|_{\Ds x_0=s/(Q^2+s)} \ .
\label{rho0}
\end{eqnarray}

\textbf{2}. In NLO we obtain \cite{BMS02,MMP02}
\begin{eqnarray}
\label{Str-T1}
  T_1(Q^2,q^2;\mu^2_{\rm F};x)
&\!\!=\!\!\!&
  C_{\rm F}\left\{ T_{\rm F}^{(1)}(Q^2,q^2;x)
  + \ln\left(\frac{\bar{Q}^2}{\mu_{\rm F}^2} \right)
  \left[T_{0} \otimes  V_+ \right](Q^2,q^2;x) \right\} \ ,
\end{eqnarray}

\begin{subequations}
\label{eq:Str-T1}
\begin{eqnarray}
\label{Def-T1F}
  T^{(1)}_{\rm F}(Q^2,q^2;x)
&\!\!\!=\!\!\!&
  T_{0}(Q^2,q^2;y)
\otimes
  \left[ {\cal T}^{(1)}(y,x) + {\rm LN}(Q^2,q^2;y)  V(y,x)_+
  \right] , \\
\label{Def-C1F}
  {\cal T}^{(1)}(x,y)
&\!\!\!=\!\!\!&
  \left[-3 V^{b}  +  g \right](x,y)_+ - 3 \delta(x-y),
\label{g}
\end{eqnarray}
where the notations of Ref.\ \cite{MMP02} and the following
abbreviation have been used:
\begin{eqnarray}
\label{Def-LN}
  {\rm LN}(Q^2,q^2;y)
&\!\!\!=\!\!\!&
    \ln\left(\frac{Q^2y+q^2\bar{y}}{\mu^2_{\rm F}}\right)
  - \ln\left(\frac{\bar{Q}^2}{\mu^2_{\rm F}}\right)\, .
\end{eqnarray}
\end{subequations}
Here
$\bar{Q}^2=-\left[(q_1-q_2)/2\right]^2=(Q^2+q^2)/2$,
and $\mu_\text{R}^2$ and $\mu_\text{F}^2$
denote, respectively, the scale of the renormalization of the
theory and the factorization scale of the process.
The kernels $V^a$ and $V^b$ are diagonal with respect to $\psi_n$
and are defined in Appendix \ref{app:Vstructure}, whereas the
kernel $g$ reads
\begin{eqnarray}
  g(x,y)
&\!\!\!=\!\!\!&
  - 2\frac{ \theta(y-x)}{y-x}\ln\left(1-\frac{x}{y}\right)
  + \left\{x \to \bar{x} \atop y \to \bar{y}  \right\}
\end{eqnarray}
and is not diagonal with respect to the Gegenbauer--polynomials.
This kernel is responsible for the apparent breaking of conformal
symmetry in the $\overline{\rm MS}$-scheme \cite{MMP02}.
Substituting (\ref{Def-T1F}) and (\ref{Def-LN}) into
Eq.\ (\ref{Str-T1}), we arrive at
\ba
\label{Str-T1-log}
  T_1(Q^2,q^2;\mu^2_{\rm F};x)
&\!\!=\!\!\!&
  T_{0}(Q^2,q^2;y)
\otimes
    \left\{C_{\rm F} {\cal T}^{(1)}(y,x)
  + {\rm Ln}(y) V^{(0)}(y,x)_+\right\} \, .
\ea
One observes that in this expression a simpler logarithm
$
 {\rm Ln}(y)
\equiv
 \ln\left[\left(Q^2y+q^2\bar{y}\right)/\mu^2_\text{F}\right]
$
appears in comparison to ${\rm LN}$ in Eq.\ (\ref{Def-LN}).
From the former expression and Eq.\ (\ref{g}), we get the following
result for the convolution
\ba
  \label{Str2-T1}
  T_1 \otimes \psi_n
=
  C_{\rm F} T_{0}(Q^2,q^2;y)
\!\!\! && \!\!\!  \otimes
  \Big\{
        -\psi_n(y)\cdot 3 \left[ 1+ v^{b}(n)\right]
        +\left[g_{+}\otimes \psi_n\right](y)
\nonumber \\
&& ~~~+\psi_n(y)\cdot  2 v(n)~{\rm Ln}(y)
  \Big\},
\ea
where the second term in Eq.\ (\ref{Str2-T1}) reads
\ba
&&
 \left[g_{+}\otimes \psi_n\right](y)
=
 \left[\frac{\pi^2}{3}-\ln^2\left(\frac{\bar{y}}{y}\right)\right]
 \psi_n(y)
\nonumber \\
&&~~~~~~~~~~~~~~~~~~~~~
  - 2\Bigg\{\int_y^1
  \left[\frac{\psi_n(x)-\psi_n(y)}{x-y}
  \right]\ln\left(1-\frac{y}{x}\right)dx
  + (y \to \bar{y})\Bigg\},
\nonumber \\
&&
  \int_y^1 \left[\frac{\psi_n(x)-\psi_n(y)}{x-y}
          \right]\ln\left(1-\frac{y}{x}\right)~dx + (y \to \bar{y})
=
  \sum^n_{l=0,2 \ldots}~G_{nl}~\psi_l(y)\, .
\label{A11}
\ea
Here we present a few partial values of the $G_{n l}$ elements:
\begin{eqnarray}
  G_{00}
& = &
    1 \, ,
\nonumber \\
  G_{2 l}
& = &
\bm{-1}  \left\{\frac{3}{2},-\frac{35}{12}\right\} \, ,
\nonumber \\
  G_{4 l}
& = &
\bm{-1}  \left\{\frac{3}{\bm{4}},~\frac{161}{72},-\frac{203}{45} \right\} \, ,
\nonumber \\
  G_{6 l}
& = &
\bm{-1}  \left\{\frac{83}{180},\frac{49}{40},~~\frac{781}{300},
  -\frac{29531}{5040}\right\} \ .
\label{A10}
\end{eqnarray}
The RHS of Eq.\ (\ref{A11}) can be further evaluated using the
following relation
\ba
&&
  2\int_0^y\frac{x^l-y^l}{y-x}\ln(1-\frac{x}{y})~dx
=
  y^l\Big[(\psi(l+1)-\psi(1))^2-(\psi^{(1)}(l+1)
  -\psi^{(1)}(1))\Big] \, .~~~~~
\ea
The discontinuity of $T_1 \otimes \psi_n$ in \textbf{Eq.\ (\ref{Str2-T1})}
consists of one part, determined by the discontinuity of $T_0$ and following
directly from Eq.\ (\ref{rho0}), and a second nontrivial part entailed
by the logarithm, as one can also verify from the analysis in
\cite{SchmYa99}.
Then one has
\ba
& &
  \frac{\mathbf{Im}}{\pi}  \left\{
   T_0(Q^2,-s;y)\otimes
   \ln\left(\frac{(Q^2+s)y-s}{\mu^2_{\rm F}} \right)
     \right. \nonumber \\
& = &\left.
  \frac1{\bm{(Q^2+s)}} \int^1_0\frac{d y}{y-y_0}
  \left[\left(\ln(y-y_0) - \ln(y_0)\right)+
  \ln\left(\frac{s}{\mu^2_{\rm F}}\right) \right]
  \right\}
\nonumber \\
& = &
 \frac{1}{\bm{(Q^2+s)}} \int^1_0 d y\left[\delta(y-y_0)
             \ln\left(\frac{s}{\mu^2_{\rm F}}\right)
 -\left(\frac{\theta(y_0>y)}{y-y_0}\right)_{+(y_0)}\right]
 \ ,
\label{A12}
\ea
\textbf{where  $\Ds y_0=\bm{s/(Q^2+s)}$.}
Substituting Eq.\ (\ref{A11}) into (\ref{A10}) by taking into account
Eqs.\ (\ref{rho0}) and (\ref{A12}),
and collecting all terms, one finally arrives for the spectral
density $\bar{\rho}^{(1)}_n(y)$ at the final results given by
Eqs.\ (\ref{eq:rho1a})--(\ref{eq:rho1b}).
Note that the last term in Eq.\ (\ref{A12}) generates the term
(\ref{eq:rho1c}).
To obtain the ``algebraic'' form (\ref{eq:rho1-n}), one should insert
the relation (\ref{4.3}) into the LHS of Eq.\ (\ref{4.b}).
The first partial $b_{nl}$ coefficients of the expansion of the final
result over the Gegenbauer harmonics are determined to be
\ba
  b_{00}
& = &
  \frac{3}{2}\, ,
\nonumber \\
  b_{2 l}
& = &
  \left\{\frac{3}{2},-\frac{5}{2},\frac{25}{12}\right\}\, ,
\nonumber \\
  b_{4 l}
& = &
  \left\{\frac{3}{2},-\frac{5}{4},\frac{7}{12},-\frac{4}{9},\frac{49}{20} \right\}\, ,
\nonumber \\
  b_{6 l}
& = &
  \left\{\frac{3}{2},-\frac{31}{30},\frac{7}{12},-\frac{19}{20},~\frac{11}{30},-\frac{13}{6},-\frac{761}{280}
  \right\}\, \label{A17}\\
  \ldots \nonumber
\ea

\textbf{3}. In NNLO we extract the $b_0$-proportional contribution
by collecting all terms in the general structure of $T_{2}$
computed in \cite{MMP02}.
The result is
\begin{eqnarray}
\label{Str-T2}
&&
  T_{2}(Q^2,q^2;\mu^2_{\rm F};\mu^2_{\rm R};x) \to  b_0 C_{\rm F}\cdot
  \Big\{ T_\beta^{(2)}(\omega,x)
  + \ln\left(\frac{\bar{Q}^2}{\mu_{\rm F}^2}\right)
  \left[T_{0}\otimes (V^{(1)}_{\beta})_+ \right](\omega,x)
\nonumber \\
&&
 - \ln\left(\frac{\bar{Q}^2}{\mu_{\rm R}^2}\right)
   T_{1}(Q^2,q^2;\mu^2_{\rm F};x)\frac{1}{C_{\rm F}}+\frac{1}{2}
   \ln^2\left(\frac{\bar{Q}^2}{\mu_{\rm F}^2}\right)
   \left[T_{0} \otimes   V^{}_+\right](\omega,x)\Big\} \, ,
\end{eqnarray}
where
\begin{eqnarray}
&&
  T^{(2)}_\beta(\omega,x)
=
  T_{0}(Q^2,q^2;y)\otimes
\nonumber \\
&&
  \Bigg\{{\cal T}^{(2)}_{\beta}(y,x) + {\rm LN}(\omega,y)
  \left[( V^{(1)}_{\beta})_+ - {\cal T}^{(1)}\right](y,x)
  -\frac{1}{2} {\rm LN}^2(\omega,y) V(y,x)_+
  \Bigg\} \, .
\label{Def-T2b}
\end{eqnarray}
We use here for the elements of $T_{2}$ the notation
$T^{(2)}_\beta,~{\cal T}^{(2)}_{\beta},~{\cal T}^{(1)}$
and $V^{(1)}_{\beta}$, introduced in \cite{MMP02}, that differs
from the previous one by a factor of 2.
We have
\begin{eqnarray}
\label{Def-C1Fnew}
  {\cal T}^{(2)}_{\beta}(x,y)
&\!\!\!=\!\!\!&
  \Bigg[\frac{29}{12} V^{a} + \dot{V}^{a}
        - \frac{209}{36} V  - \frac{7}{3} \dot{V}
        - \frac{1}{4} \ddot{V} + \frac{19}{6} g
        + \dot{g}
  \Bigg]_+\!\!(x, y)  - 6  \delta(x-y)\, ,~~~~
\end{eqnarray}
while ${\cal T}^{(1)}$ is defined in (\ref{g}).
One observes that these $\cal T$-kernels are calculated in terms of
the kernels $V$ and $g$, which enter the evolution kernel $V^{(1)}$,
and the derivatives $\dot{V}, \ddot{V}$ that will be discussed
in Appendix~\ref{app:Vstructure}.
The origin of $\dot{g}$ has been clarified in Sec.\ 3 of
Ref.\ \cite{MMP02}.

Under the assumption that $T_{2} \to  b_0 \cdot T_{\beta}$,
it turns out that the entire ${\rm LN}(\omega,y)$ dependence
appears only inside the term
$\Ds {\rm Ln}(y)$ [cf.\ Eq.(\ref{eq:L-n})],
in analogy to the NLO case, see (\ref{Str-T1-log}):
\begin{subequations}
\begin{eqnarray}
&&
  T_{2} \to  b_0 \cdot T_{\beta}; \\
&&
  T_{\beta}
=
  C_{\rm F} T_{0}(Q^2,q^2;y)\otimes
  \Bigg\{\!\! -\ln\left(\frac{\mu_{\rm F}^2}{\mu_{\rm R}^2}\right)
  \left[{\cal T}^{(1)}(y,x) +
  {\rm Ln}(y)  V^{}_+(y,x)\right]
\nonumber  \\
&&
  \hspace*{4.52cm} + \ {\cal T}^{(2)}_{\beta}(y,x)
  +{\rm Ln}(y) \left[(V^{(1)}_{\beta})_+
  -{\cal T}^{(1)}\right](y,x)
\nonumber \\
&&
  \hspace*{4.52cm} - \ \frac{1}{2}~{\rm Ln}^2(y)
  V(y,x)_+ \Bigg\}\, .
\label{T2-final-log2}
\end{eqnarray}
\end{subequations}
This important property of the $T_{\beta}$ structure has already been
mentioned in \cite{MMP02}.
While all terms in (\ref{T2-final-log2}) contribute to the
discontinuity of $T_{\beta}$, only the last term, which contains the
square of a logarithmic expression, contributes a new type of
discontinuity.
On the other hand, the term proportional to
$\ln(\mu_{\rm F}^2/\mu_{\rm R}^2)$
equals $T_1$, as it can be seen from Eq.\ (\ref{Str-T1-log}).
Thus, the final structure of $T_{\beta}$ assumes the form
\begin{eqnarray}
&&
  T_{\beta}
= -\ln\left(\frac{\mu_{\rm F}^2}{\mu_{\rm R}^2}\right)T_{1}
 + ~C_{\rm F} T_{0}(Q^2,q^2;y)\otimes
  \Big\{\nonumber \\
&&
  {\cal T}^{(2)}_{\beta}\bm{(y,x)} +
  {\rm Ln}(y) \left[(V^{(1)}_{\beta})_+
  -{\cal T}^{(1)}\right](y,x)
  - \frac{1}{2}{\rm Ln}^2(y)
  \bm{V_+(y,x)} \Big\}\, .
\label{eq:T-beta-2}
\end{eqnarray}
From this expression, one sees that the first term, which contains no
logarithm at all, has no influence on the discontinuity of $T_0$.
On the other hand, the remaining terms contain logarithms, which
do affect the discontinuity, with the last one being the new
contribution first appearing at the NNLO level.

Then, the partial amplitude $T_{\beta}\otimes \psi_n$ reads
\begin{eqnarray}
  T_{\beta}\otimes \psi_n
&=&
  -\ln\left(\frac{\mu_{\rm F}^2}{\mu_R^2}\right)T_{1}\otimes \psi_n
\nonumber \\
&&
  +C_{\rm F} T_{0}\otimes
  \Big\{
        {\cal T}^{(2)}_{\beta} +
        {\rm Ln}\cdot \left[(V^{(1)}_{\beta})_+
        -{\cal T}^{(1)}\right]
        - \frac{1}{2}{\rm Ln}^2
        \cdot (2 v(n))
  \Big\} \otimes \psi_n \, .
\label{T2beta-n}
\end{eqnarray}

Let us now calculate the discontinuity entailed by the new contribution
in NNLO,
\ba
&&\frac{ \mathbf{Im}}{\pi} \left\{
   T_0(Q^2,-s;y)\otimes
   \ln^2\left(\frac{(Q^2+s)y-s}{\mu^2_{\rm F}} \right)
=
\right. \nonumber \\
&& \left.
  \frac1{\bm{(s+Q^2)}}\cdot \int^1_0\frac{d y}{y-y_0}
  \left[\left(\ln(y-y_0) - \ln(y_0)\right)+
  \ln\left(\frac{s}{\mu^2_{\rm F}}\right) \right]^2
 \right\} \ .
\label{eq:appL2}
\ea
Substituting identity \cite{SchmYa99}
\ba
  &&\frac1{\pi} \mathbf{Im}\left[\frac{\ln^2(y-y_0)}{y-y_0}\right]
=
  \delta(y-y_0)\left[\ln^2(y_0)-\pi^2/3\right]
  -2\left[ \theta(y_0>y)\frac{\bm{\ln(y_0-y)}}{y-y_0}\right]_+ ~~~
\ea
into Eq.\ (\ref{eq:appL2}), we obtain for its RHS
\begin{eqnarray}
  \frac{1}{\bm{(Q^2+s)}} \int^1_0 d y \!\!\!
&&
  \left( \delta(y-y_0)
  \left[\ln^2\left(\frac{s}{\mu^2_{\rm F}}\right)-\pi^2/3\right]
\right.\nonumber \\
&&\left. - 2
  \left\{\frac{\theta(y_0>y)}{y-y_0}\left[\ln(|y-y_0|) - \ln(y_0)
  + \ln\left(\frac{s}{\mu^2_{\rm F}}\right)\right]
  \right\}_{\bm{+(y_0)}} \ \right) \ .
\nonumber \\
  &&
\end{eqnarray}

\section{Structure of the evolution kernel $V$}
\renewcommand{\theequation}{\thesection.\arabic{equation}}
\label{app:Vstructure}\setcounter{equation}{0}
This Appendix compiles the crucial properties of the evolution
kernels borrowing results from \cite{MS98b,MS00} and \cite{MMP02}.
Following \cite{MR86ev,MV08}, we introduce the auxiliary kernels
$V^{a,b}(x,y;\lambda)$, viz.,
\ba
  V_{+}^{a}(x,y;\lambda)
&=&
  \left[
        \theta(y>x)\left(\frac{x}{y}\right)^{1+\lambda}
 +\left\{x \to \bar{x} \atop y \to \bar{y}\right\}
  \right]_+ \, ,
\label{Def-va}
\\
  V_{+}^{b}(x,y;\lambda)
&=&
  \left[\frac{\theta(y>x)}{y-x}\left(\frac{x}{y}\right)^{1+\lambda}
 +\left\{x \to \bar{x} \atop y \to \bar{y}\right\}
  \right]_+ \, ,
\label{Def-vb}
\ea
and their sum
$V(x,y;\lambda)=V_{+}^{a}(x,y;\lambda)+V_{+}^{b}(x,y;\lambda)$
that includes the main of the logarithmic contributions---generated
by the one-loop renormalization of the running coupling---accompanied
by the factor $a_s\ln(x/y)$.
Just the effects of this renormalization lead to those contributions
that are proportional to $b_0$ and pertain to the derivatives of the
auxiliary kernel
$V(x,y;\lambda)$, i.e.,
\ba
 \dot{V}_+(x,y)
\equiv
  2V'(x,y;\lambda)\mid_{\lambda=0}
=
  2 \left[\theta(y>x)\frac{x}{y}
  \left(1+\frac{1}{y-x}\right)\ln\left(\frac{x}{y}\right)
  + \left\{x \to \bar{x} \atop y \to \bar{y}  \right\}\right]_+ ,
\label{Vdot} \\
  \ddot{V}_+(x,y)
\equiv
 2V''(x,y;\lambda)\mid_{\lambda=0}
  =
  2 \left[\theta(y>x)\frac{x}{y}\left(1+\frac{1}{y-x}\right)
  \ln^2\left(\frac{x}{y}\right)
  + \left\{x \to \bar{x} \atop y \to \bar{y}  \right\}\right]_+ \, ,
\label{Vddot}
\ea
we already faced in Eq.\, (\ref{Def-C1Fnew}).
The $b_0$-proportional part $V^{(1)}_{\beta}$ of the NLO kernel
$V^{(1)}$, that enters $T_\beta$ in Eqs.\, (\ref{T2-final-log2}) and
(\ref{eq:T-beta-2}) for $T_{\beta}$, contains the non-logarithmic terms
\ba
 V^{(1)}_{\beta}= \dot{V} +\frac{5}{3}~V + \bm{2}V^{a}\ .
\label{Vbeta}
\ea
These contributions can be obtained from the generating kernel
\cite{MS00}
\ba
\label{eq:Vgeneral}
  V_{\beta}(x,y|\lambda)
=
 \bm{2} \left[(1+\lambda)V^{a}(x,y;\lambda)+ V^{b}(x,y;\lambda)
  \right]_{+}C(\lambda)
\ea
for any order of $b_0$.
Here $C(\lambda)$ is an analytic function in the variable
$\lambda$ with $C(0)=1$.
To obtain the $b_0$-contribution at any desired fixed order of
the parameter $a_s b_0$, one has to expand the kernel
$V_{\beta}(x,y|a_s b_0)$ in a Taylor series with respect to $a_s$ up
to this order.
Hence $V_{\beta}(x,y|0)=V_+(x,y)$, while the first differentiation of
$V_{\beta}$ with respect to $a_s$,
\ba
\label{V-gen}
  \frac{d}{d a_s}V_{\beta}(x,y|a_s b_0)\Big|_{a_s=0}
=
  b_0\cdot V^{(1)}_{\beta}(x,y)\, ,
\ea
leads to $V^{(1)}_{\beta}$.
The generalized kernel $V_{\beta}(x,y|\lambda)$ has been derived from
the diagrams for the ordinary one-loop kernels, generated by replacing
single gluon lines by a sum of renormalon-chain insertions.
The coefficient $C(\lambda)$ in Eq.\ (\ref{V-gen}) accumulates
non-logarithmic parts of these renormalon-chain contributions and
was determined in \cite{MS98b,MS00}.

\section{Main elements of the NNLO partial amplitude}
\renewcommand{\theequation}{\thesection.\arabic{equation}}
\label{App-NNLOelements}\setcounter{equation}{0}
We present now the main elements of the partial amplitudes
$T_{\beta}\otimes \psi_n$ and
$\left[(\bm{V^{(1)}_{\beta})_+} - {\cal T}^{(1)}\right]\otimes \psi_n$
entering Eq.\ (\ref{T2beta-n}) in Appendix \ref{app:Tstructure}.
We split each of the expressions below in two parts: the
singular part with $x\to 0$ is extracted in an explicit form
that turns out to be proportional to $\psi_n$, whereas the other part
contains an integration over longitudinal momentum fractions.

\textbf{1.} Recalling Eq.\ (\ref{g}) in App.\ \ref{app:Tstructure}, we
find

\begin{subequations}
\begin{eqnarray}
 -\left( {\cal T}^{(1)}\otimes \psi_n \right)(x)
&=&
  3 \left[ 1+ v^{b}(n)\right] \psi_n(x)+
  \left[\ln^2\left(\frac{\bar{x}}{x}\right)
  - \frac{\pi^2}{3}\right] \psi_n(x)
\nonumber \\
  && +2\left\{ \int_{\bm{x}}^1 du \left[\frac{\psi_n(u)
  -\psi_n(x)}{u-x}\right]\ln\left(1-\frac{x}{u}\right)
   +   (x \to \bar{x}) \right\} ,
\label{C1} \\
-\left( {\cal T}^{(1)}\otimes \psi_0 \right)(x)
&=&
\left[3+ \ln^2\left(\frac{\bar{x}}{x}\right)
- \frac{\pi^2}{3}+2 \right] \psi_0(x) \ ,
\label{C2}
\end{eqnarray}
\end{subequations}
\begin{subequations}
\label{eqCapp:V}
\begin{eqnarray}
  - \frac{1}{2}{\rm Ln}^2
  \cdot V_+\otimes \psi_n
&=&
  - \frac{1}{2}{\rm Ln}^2 \cdot \bm{2}v^{}(n)\cdot \psi_n \, ,
\label{C3} \\
  - \frac{1}{2}{\rm Ln}^2
\cdot V_+\otimes \psi_0 &=& 0 \, .
\label{C4}
\end{eqnarray}
\end{subequations}

\textbf{2.} According to the definition of $V^{(1)}_{\beta}$ in
(\ref{Vbeta}) and the definitions of $\dot{V}, \dot{V}^{a} $ in
(\ref{Vdot}), we have
\begin{subequations}
\label{eqCapp:Vot}
\ba
  \left(\dot{V}^{}_+\otimes \psi_n \right)(x)
&=&
 \bm{2} \left\{\int^{1}_{x} \frac{x}{u}\left(1+\frac{1}{u-x}\right)
  \ln\left(\frac{x}{u}\right)\left[\psi_n(u)-\psi_n(x)\right]du
  +(x \to \bar{x}) \right\}
\nonumber \\
&&
  - \left[ -\bar{x}\ln^2(x)-x\ln^2(\bar{x})+\frac{7}{2}
  +\ln^2\left(\frac{\bar{x}}{x}\right)-\frac{\pi^2}{3}\right]
  \psi_n(x) \, ,
\label{C5} \\
  \left(\dot{V}^{}_+\otimes \psi_0 \right)(x)
&=&
  -\left\{6\bar{x} \ln(\bar{x})+6x \ln(x)+
  \left[5+ \ln^2\left(\frac{\bar{x}}{x}\right)
  - \frac{\pi^2}{3}\right]\psi_0(x) \right\}\, ,
\label{C6}
\ea
\end{subequations}
\begin{subequations}
\label{eqCapp:Vbeta}
\ba
  \left[ (V^{(1)}_{\beta})_+ \otimes \psi_n \right](x)
&=&
  \left(\dot{V}_+\otimes \psi_n\right)(x)
  + \left[ \frac{5}{3}\bm{2} v(n) + \bm{2}v^{a}(n)\right]\psi_n(x), \\
  \left[(V^{(1)}_{\beta})_+ \otimes \psi_0 \right](x)
&=&
  \left(\dot{V}_+\otimes \psi_0 \right)(x)
\nonumber \\
&=&
  -\Big\{6\bar{x} \ln(\bar{x})+6x\ln(x)+
  \left[5+ \ln^2\left(\frac{\bar{x}}{x}\right)- \frac{\pi^2}{3}\right]
  \psi_0(x) \Big\},
\label{C8}
\ea
\end{subequations}
\begin{subequations}
\label{eqCapp:Vadot}
\begin{eqnarray}
  \dot{V}^{a}_+\otimes \psi_n
&=&
  \left\{\int^{1}_{x} \frac{x}{u}\ln\left(\frac{x}{u}\right)
  [\psi_n(u)-\psi_n(x)]du +
   x \left(\frac{1}{\bm{4}} - \frac{\ln^2(x)}{\bm{2}}\right)\psi_n(x) \right\}
\nonumber \\
&& ~~~~
  +(x \to \bar{x}) \ ,
\label{C9} \\
  \dot{V}^{a}_+\otimes \psi_0
&=&
  3 \left(\bar{x}\ln(\bar{x})+ x\ln(x)+3x\bar{x} \right)\, .
\label{C10}
\end{eqnarray}
\end{subequations}
Using the expressions obtained in Eqs.\ (\ref{C2}) and (\ref{C8}),
we obtain
 \begin{eqnarray}
 \left[(\bm{V^{(1)}}_{\beta})_+ - {\cal T}^{(1)}\right]\otimes \psi_0
=
  -6\bar{x} \ln(\bar{x})-6x \ln(x) \, . \label{C11}
\end{eqnarray}

\textbf{3.} The most characteristic contribution to $T_{\beta}$ is
represented by the term ${\cal T}^{(2)}_{\beta}$.
The result for the convolution is \cite{MMP02}
\begin{eqnarray}
\label{}
  {\cal T}^{(2)}_{\beta}\otimes \psi_n
&=&
   \Bigg[ \frac{29}{12}  \bm{2}v^{a}(n) \psi_n
  - \frac{209}{36} \bm{2}v(n) \psi_n + \bm{2}\dot{V}^{a}_+\otimes \psi_n
  -\frac{7}{3} \dot{V}_+\otimes \psi_n
  - \frac{1}{4} \ddot{V}_+\otimes \psi_n
\nonumber \\
&&
  ~~~ + \frac{19}{6} g_+\otimes \psi_n
  + \dot{g}_+\otimes \psi_n \Bigg](x)  - 6\psi_n(x) \, .
\label{C12}
\end{eqnarray}
This expression contains a couple of new kernel elements,
notably, $\ddot{V}_+$ and $\dot{g}_+$.
The convolution expressions for each of these kernels with the
Gegenbauer harmonics are displayed below, starting with the general
case $n$, and followed by the zeroth-order harmonic.
\begin{subequations}
\label{eqCapp:V2dot}
\begin{eqnarray}
  \left(\ddot{V}^{}_+\otimes \psi_n \right)(x)
&=&
 \bm{2} \left[\int^{1}_{x} \frac{x}{u}\left(1+\frac{1}{u-x}\right)
  \ln^2\left(\frac{x}{u}\right)
  [\psi_n(u)-\psi_n(x)]du +(x \to \bar{x})\right]
\nonumber \\
&&
 ~~ + \psi_n(x)\left[
                     4\ln(x){\rm Li_2}(x)-4{\rm Li_3}(x)
                     -\frac2{3}x\ln^3(x)+2\ln(\bar{x})\ln^2(x)
                     \right.
\nonumber \\
&& \left. ~~~~~~~~~~~~~~
                     + 4-\frac{x}{2}\phantom{\frac2{3}}\!\!\!
               \right]
  + (x \to \bar{x}) \, ,\\
  \left(\ddot{V}^{}_+\otimes \psi_0 \right)(x)
&=&
  6x \bar{x}
  \Big\{
        4\left[
               \ln(\bar{x}){\rm Li_2}(\bar{x})
               +\ln(x){\rm Li_2}(x)- {\rm Li_3}(x)- {\rm Li_3}(\bar{x})
         \right]
\nonumber\\
&&
~~~~~~ - \frac{2}{3}(\ln^3(x)+\ln^3(\bar{x}))+ 2\ln(x)\ln(\bar{x})
       (\ln(x)+\ln(\bar{x})) + 6
  \Big\}
\nonumber\\
&&
   -6x\left(\ln^2(x)+\ln(x)\right) -6\bar{x}\left(\ln^2(\bar{x})
   +\ln(\bar{x})\right)\, ,
\label{C13}
\end{eqnarray}
\end{subequations}
\begin{eqnarray}
  \dot{g}_{+}(x,y)
&=&
   \bm{2} \frac{ \theta(y-x)}{y-x}
  \left[
        {\rm Li}_2\left(1-\frac{x}{y}\right)- {\rm Li}_2(1)
        -\frac{1}{2}\ln^2\left(1-\frac{x}{y}\right)
  \right] + \left\{x \to \bar{x} \atop y \to \bar{y}  \right\},
\label{dotg+}
\end{eqnarray}
\begin{subequations}
\label{eqCapp:gdot}
\begin{eqnarray}
  \left[\dot{g}_{+}\otimes \psi_n\right](x)
&=&
  \bm{2}\Bigg\{\int_x^1 \left[\frac{\psi_n(u)-\psi_n(x)}{u-x}\right]
  \left[
        {\rm Li}_2(1-\frac{x}{u})- {\rm Li}_2(1)-\frac{1}{2}
        \ln^2\left(1-\frac{x}{u}\right)
  \right] du
\nonumber \\
&&
  +(x \to \bar{x}) \Bigg\}
  + \psi_n(x) \left[
                    \ln(x)\ln^2(\bar{x})
                    -\frac{1}{3}\ln^3(\bar{x})
                    -\frac{\pi^2}{3}\ln(\bar{x})
                    +4 \text{Li}_3(x) \right.
\nonumber \\
&&  \left.
  ~~~~~~~~~~~~~~~~~~~~~~~~~~~~ -6 \zeta (3)
              \phantom{\frac1{3}}\!\!\!\right]
  + (x \to \bar{x}) \, ,
\label{dotg+psin} \\
  \left[\dot{g}_{+}\otimes \psi_0\right](x)
&=&
  6x\bar{x}\Big\{\ln(x)\ln^2(\bar{x})+\ln^2(x) \ln(\bar{x})
  -\frac1{3}\left[\ln^3(\bar{x})+\ln^3(x)\right]
\nonumber \\
&&
  ~~~~~~ -\frac{\pi^2}{3} \left[\ln(\bar{x})+\ln(x)\right]+
  4 \left[\text{Li}_3(x)+\text{Li}_3(\bar{x})\right]
  -12\zeta(3)-2 \Big\}
\nonumber \\
&&
  -12 \bar{x} \ln(\bar{x})-12 x\ln(x) \ .
\label{C16}
\end{eqnarray}
\end{subequations}
Finally, substituting Eqs.\ (\ref{A11}), (\ref{C6}), (\ref{C10}),
(\ref{C13}), and (\ref{C16}) into Eq.\ (\ref{C12}) for $n=0$,
and taking into account that $v^a(0)=v(0)=0$, one obtains for
${\cal T}^{(2)}_{\beta}\otimes \psi_0$ the expression
\begin{eqnarray}
  \label{}
  {\cal T}^{(2)}_{\beta}\otimes \psi_0
&=&
  x\bar{x}\Big\{30\left[\text{Li}_3(x)+\text{Li}_3(\bar{x})\right]-
  6\left[ \ln(\bar{x}){\rm Li_2}(\bar{x})+\ln(x){\rm Li_2}(x)\right]
  -\left[\ln^3(\bar{x})+\ln^3(x)\right]
\nonumber \\
&&
  -5\ln^2\left(\frac{\bar{x}}{x}\right)+
  \left[\ln(\bar{x})+\ln(x)\right]
  \left(3\ln(\bar{x})\ln(x)-2\pi^2\right)-72\zeta(3)\bm{+\frac{5}{3}\pi^2}-7
  \Big\}
\nonumber \\
&&
  + \frac{19}{2}\left[\bar{x}\ln(\bar{x})+x\ln(x)\right]+
  \frac{3}{2}\left[\bar{x}\ln^2(\bar{x})+x\ln^2(x)\right]\, .
\label{C18}
\end{eqnarray}

\end{appendix}

\end{document}